\documentclass[10pt,twocolumn]{article}

\usepackage[margin=0.85in]{geometry}
\usepackage{microtype}
\usepackage{booktabs}
\usepackage{array}
\usepackage{tabularx}
\usepackage{longtable}
\usepackage{graphicx}
\usepackage{subcaption}
\usepackage{amsmath}
\usepackage{amssymb}
\usepackage{xurl}
\usepackage{hyperref}
\usepackage[nameinlink,noabbrev]{cleveref}
\usepackage{balance}
\usepackage{enumitem}

\graphicspath{{figures/}}
\hypersetup{breaklinks=true}

\title{SCION: Size-aware Policy Orchestration for Nonstationary Object Caches (Long Paper Version)}
\author{
  Qizhi Wang\\
  PingCAP, Data \& AI-Innovation Lab\\
  Beijing, China\\
  \texttt{qizhi.wang@pingcap.com}
}
\date{}

\begin{document}
\maketitle

\begin{abstract}
Object caches underpin modern cloud and edge services (CDNs, object stores, and large-scale web/KV caching),
yet production workloads are heterogeneous, nonstationary, and throughput-constrained.
Recent strong non-ML policies (e.g., SIEVE and S3-FIFO) raise the bar for any learned approach:
added intelligence must be overhead-aware, robust under drift, and competitive with strong baselines.
We present \textsc{SCION}, a lightweight \emph{policy-orchestration} framework that selects among a small set of deployable cache policies
using a tiny workload fingerprint computed off the critical path.
Our prototype instance, \textsc{AUTO}, computes a short-prefix fingerprint
(log p50/p90/mean object size, log tail ratio, cacheable fraction, unique ratio, and log cache size; default $N{=}200$k)
and uses a linear selector trained offline with leave-one-trace-out validation to choose among
GDSF, S3-FIFO, SIEVE, LHD, W-TinyLFU-AV, and DynamicAdaptiveClimb; a simpler \textsc{SCION-P90} variant uses only a p90 threshold.
Using a CPU-only, trace-driven evaluation built on \texttt{libCacheSim} and 30 public object-cache traces from \texttt{cache\_dataset}
(5M requests each, plus 20M for four long traces), and a separate HR-Cache (hazard-rate cache) simulator on a 10-trace subset,
\textsc{AUTO} improves cacheable-only object miss ratio (OMR) over SIEVE on a majority of workloads,
stays close to the best single expert on average (GDSF remains the best average-OMR baseline), and enables explicit OMR/BMR tradeoff selection
while remaining competitive on byte miss ratio (BMR).
Under a fast-policy budget, \textsc{AUTO-fast} achieves a lower $\alpha{=}0.5$ cost than the best fixed fast policy.
We quantify prefix-selection stability over post-prefix windows, fingerprint cost, and per-policy throughput overheads, and release a fully reproducible benchmark pipeline.
SCION operationalizes a learning-lite orchestration design point that reduces regime-mismatch risk while keeping the hot path unchanged.
\end{abstract}

\section{Introduction}
\label{sec:intro}
Object caches are ubiquitous in future-generation computing systems (CDNs, cloud object stores, web caches, and key--value (KV) caching layers)
and operate under strict throughput and latency constraints.
Unlike CPU caches, object sizes are heavy-tailed and misses have non-uniform costs; a miss on a large object can dominate backend bandwidth,
tail latency, and cross-region traffic. As a result, byte miss ratio (BMR) is often a primary operator metric alongside OMR.
Workloads also drift with diurnal patterns, multi-tenancy, and shifting content mixes.
Recent systems work has introduced simple yet strong policies, notably SIEVE and S3-FIFO, that raise the baseline for any learned approach~\cite{sieve,s3fifo}.

\paragraph{Problem.}
Production caches face \emph{heterogeneous} and \emph{nonstationary} workloads.
Even strong policies exhibit regime-specific weaknesses: a policy that excels for KB-scale objects can degrade under MB-scale objects (and vice versa).
This makes ``pick one policy'' risky---in our traces, the wrong choice can increase cacheable-only OMR by $>0.2$ on some workloads---but also suggests a pragmatic opportunity:
if workloads can be fingerprinted cheaply, then \emph{choosing the right strong policy} can outperform any fixed choice without heavy per-request ML.

\paragraph{Motivation.}
Operators rarely deploy heavyweight ML on the hot path, but they do switch policies across tiers or services.
The gap is \emph{automation with low overhead}: a selector that is cheap enough to run outside the per-request path,
yet accurate enough to avoid catastrophic regime mismatch.
\textsc{SCION} targets this point in the design space.

\begin{figure*}[t]
  \centering
  \includegraphics[width=0.92\textwidth]{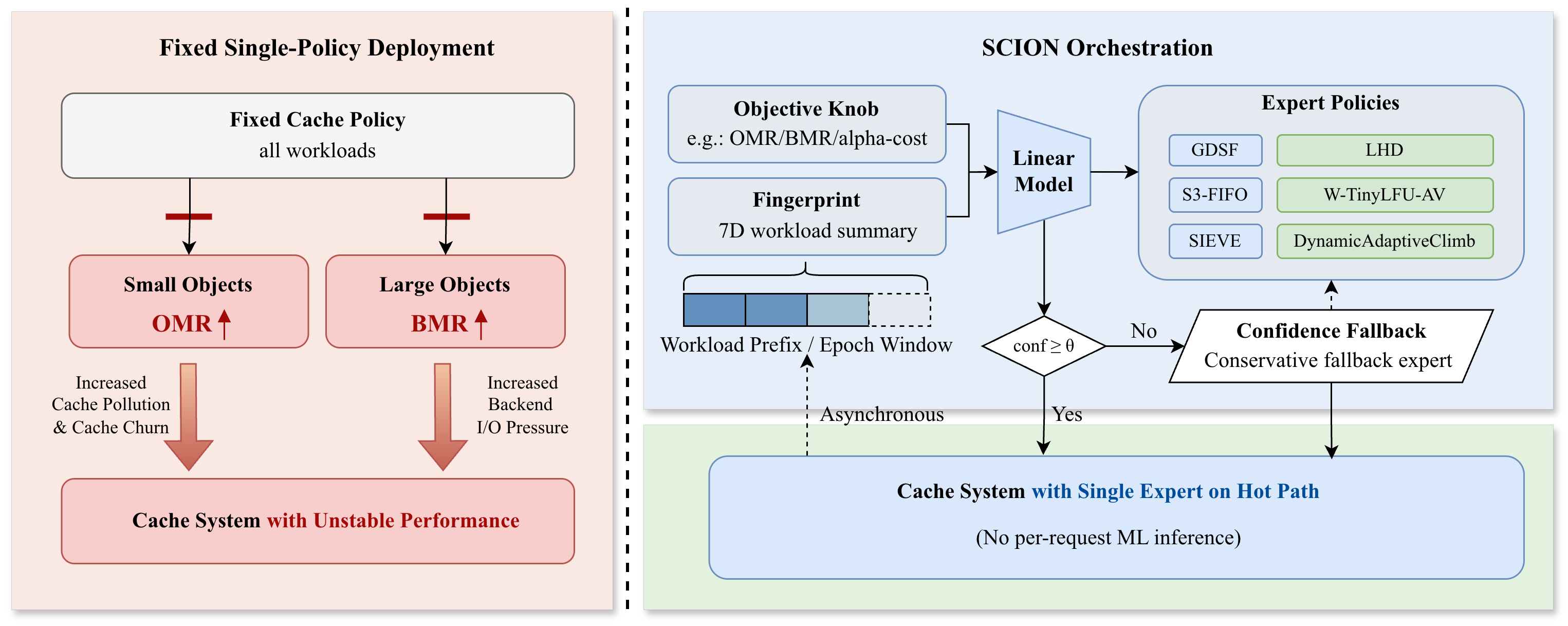}
  \caption{SCION architecture. A lightweight off-path selector chooses one expert from a small portfolio using a short-prefix workload fingerprint, while the request path remains a single active expert cache with standard hit/miss handling.}
  \label{fig:scion_arch}
\end{figure*}

\paragraph{Thesis and approach.}
The current frontier in learned caching is moving away from heavy per-request inference toward minimal-overhead, robust, deployable designs~\cite{lrb,cacheus,halp,mat,threeLcache}.
We adopt this perspective and propose \textsc{SCION}, a learning-lite orchestration layer:
use a tiny fingerprint (computed rarely) to select among a small set of expert policies (implemented with existing, battle-tested logic).
The serving path remains purely ``systems code''; learning/inference is off the critical path.
When the selector is uncertain, we fall back to a conservative expert (GDSF by default for OMR), and quantify fallback sensitivity in \Cref{tab:fallback_sensitivity}.
Figure~\ref{fig:scion_arch} summarizes the deployment model: a lightweight off-path selector chooses one expert from a small portfolio using a short-prefix workload fingerprint, while the request path remains a single active expert cache.

\paragraph{Terminology and notation.}
We consider a cache of capacity $C$ bytes serving requests for objects of size $s$.
The \emph{per-request path} is the hot path that executes admission and eviction for each request.
\emph{Off-critical-path inference} refers to computations performed outside that path (e.g., once per $N$-request epoch).
We report object miss ratio (OMR) and byte miss ratio (BMR); \emph{cacheable-only} metrics restrict to requests with $s \le C$.
$N$ denotes the prefix/window length used for fingerprinting; $\tau$ is the p90 threshold in \textsc{SCION-P90};
p90 is the 90th percentile of object size in the prefix.
Let $x \in \mathbb{R}^d$ be the fingerprint feature vector with $d{=}7$ features
(log p50/p90/mean size, log tail ratio, cacheable fraction, unique ratio, log cache size),
where the \emph{unique ratio} is the fraction of distinct object IDs in the prefix ($1{-}$reuse).
\textsc{SCION} is the framework; \textsc{AUTO} is its default learned selector instance.

\paragraph{Contributions.}
\begin{itemize}[leftmargin=*]
  \item \textbf{Learning-lite orchestration.} A lightweight policy-orchestration abstraction with a reject-option fallback: small action space, cheap fingerprints, and strong baselines as experts (\cref{sec:method}).
  \item \textbf{A minimal instance.} \textsc{AUTO} uses a tiny multi-feature fingerprint (log p50/p90/mean size, log tail ratio, cacheability, unique ratio, cache size) and a linear selector (trained via leave-one-trace-out) to choose among six strong experts; a simple \textsc{SCION-P90} threshold is an ablation (\cref{sec:auto}).
  \item \textbf{Reproducible evaluation.} A CPU-only, trace-driven benchmark and scripts based on \texttt{libCacheSim}~\cite{libcachesim} and 30 public traces from \texttt{cache\_dataset}~\cite{cache_dataset}, with bootstrap confidence intervals and trace integrity diagnostics (\cref{sec:eval}).
  \item \textbf{Robustness analysis.} Sensitivity studies over $(N,\tau)$ and fingerprints, prefix-stability analysis over post-prefix windows, out-of-distribution generalization across dataset families, spliced-trace nonstationarity tests with hysteresis and bandit baselines, plus worst-5\% window tail analysis (\cref{sec:results}).
  \item \textbf{Overhead accounting.} Microbenchmarks for fingerprint cost and per-policy throughput, plus a separate HR-Cache baseline evaluated on an overlapping subset (\cref{sec:results}).
  \item \textbf{Engineering and artifact.} We integrate DynamicAdaptiveClimb into \texttt{libCacheSim}, build a trace-conversion and evaluation pipeline for HR-Cache, and provide end-to-end scripts to download traces, run all experiments, and regenerate tables/figures.
  We surface key per-workload wins and CIs in the paper, while full per-workload, spliced-trace, and tradeoff example tables are included in the artifact to keep the paper concise.
  \item \textbf{Operational relevance.} \textsc{AUTO} improves cacheable-only OMR over SIEVE on most workloads, matches or exceeds the best expert on small-object regimes, and supports explicit OMR/BMR tradeoffs useful to cloud and edge operators (\cref{sec:results}).
\end{itemize}

\section{Background and Motivation}
\label{sec:bg}
\paragraph{Object caching and size effects.}
Object caches must decide (i) whether to admit an object on a miss and (ii) what to evict when space is needed.
Because objects vary widely in size, a policy that ignores size may suffer from pollution and poor byte efficiency.
Classic approaches include TinyLFU-style admission~\cite{tinylfu} and size-aware scoring such as the GreedyDual-Size family (GDS/GDSF)~\cite{cao1997costaware}.

\paragraph{The new baseline reality.}
SIEVE and S3-FIFO show that simple mechanisms can be highly competitive in object caches~\cite{sieve,s3fifo}.
This shifts what is publishable: ``ML beats LRU'' is no longer interesting; the bar is beating strong non-ML baselines \emph{and} accounting for overhead and robustness.

\paragraph{Why orchestration.}
Many deployments already swap policies across tiers or services; however, manual selection is brittle under workload drift.
An orchestration layer can automate selection while keeping the per-request path simple.
This idea echoes mixture-of-experts in caching (e.g., Cacheus)~\cite{cacheus} and minimal-overhead learned methods~\cite{mat},
and online expert tracking (e.g., LeCaR-style) but here we focus on an especially constrained, reproducible design point: a tiny fingerprint and a tiny expert set.
Learning is confined to offline selector training; at runtime, the cache executes a standard expert policy with no per-request ML.

\section{SCION Method}
\label{sec:method}
\textsc{SCION} defines the orchestration abstraction; \textsc{AUTO} is its default instantiation with a learned linear selector.
\subsection{Design principles}
\textsc{SCION} selects among experts $\mathcal{E}=\{e_1,\ldots,e_k\}$.
We design for:
\begin{itemize}[leftmargin=*]
  \item \textbf{Tiny action space.} Keep $k$ small and experts strong (e.g., SIEVE, S3-FIFO, GDSF, W-TinyLFU, LHD, AdaptiveClimb/DynamicAdaptiveClimb).
  \item \textbf{Cheap fingerprint.} Use statistics computable on a short prefix or a short window.
  \item \textbf{Off-critical-path inference.} Selection runs infrequently (per epoch/window), not per request.
  \item \textbf{Robustness-first framing.} When the selector is unsure, matching a strong baseline is acceptable; correctness is defined by avoiding catastrophic regime mismatch.
  \item \textbf{Safe fallback.} Low-confidence decisions trigger a conservative expert, and epochal selection can include hysteresis/lag to reduce oscillation.
\end{itemize}

\subsection{A minimal instance: \textsc{AUTO}}
\label{sec:auto}
Our default instance uses a tiny multi-feature fingerprint computed on the first $N$ requests of a trace or window:
log p50, log p90, log mean object size, log tail ratio (p90/p50), cacheable fraction at the target cache size,
unique ratio (distinct IDs divided by $N$), and log cache size.
The cacheable fraction is computed from the prefix alone as the fraction of prefix requests with object size $s \le C$
(no lookahead beyond the prefix).
In production, cacheable and unique ratios can be estimated with approximate distinct counters (e.g., HyperLogLog)
without changing the regime-level decision.
Let $x \in \mathbb{R}^d$ be this feature vector with $d{=}7$.
For each expert $e \in \mathcal{E}$, we learn a linear score $s_e = w_e^\top x$ and select
\[
\pi(x) = \arg\max_{e \in \mathcal{E}} s_e.
\]
\paragraph{Objective view.}
For each sample $i$ and expert $e$, let $\ell_i(e)$ be the objective value (OMR, BMR, or $\alpha$-cost),
and define regret $r_i(e)=\ell_i(e)-\min_{e'}\ell_i(e')$.
Learning a linear selector is thus a lightweight, cost-sensitive classification problem; lower misclassification implies lower expected regret,
and we report regret directly.
The confidence threshold implements a \emph{reject option}: if $\max_e p(e \mid x) < \theta$ we abstain and fall back to a conservative expert,
yielding a simple risk--coverage tradeoff that we tune on training folds.
With fallback expert $e_0$ and indicator $A{=}\mathbb{1}[\max_e p(e \mid x)\ge\theta]$, the expected regret decomposes as
$\mathbb{E}[r_i(\hat{e})A + r_i(e_0)(1-A)]$ and is bounded by
$\mathbb{E}[\Delta_i\,\mathbb{1}[\hat{e}\neq e^\star]] + \mathbb{E}[r_i(e_0)(1-A)]$,
where $\Delta_i$ is the per-sample expert spread and $e^\star$ the best expert.
We train $w_e$ with leave-one-trace-out validation on trace--size pairs; labels use full-run best-expert on the training fold only,
and all normalization/tuning is fold-local (no held-out leakage). Samples are weighted uniformly across pairs (no per-trace reweighting);
because each trace contributes the same four cache sizes after filtering, this is equivalent to per-trace weighting (artifact).
At inference time, selection is a single dot product per expert, performed once per epoch (not per request).
If the classifier confidence is low, we fall back to a conservative expert (by default, the best-average policy).
\paragraph{Confidence and fallback.}
We compute confidence as the maximum softmax probability of the linear classifier.
If $\max_e p(e \mid x) < 0.40$, we fall back to the best-average expert on the training fold for the \emph{same objective}
(unless a fixed fallback is specified), so BMR-centric selectors default to a BMR-optimized fallback.
\Cref{tab:fallback_sensitivity} shows that, on this suite, GDSF minimizes average and tail regret for both OMR and $\alpha{=}0.2$
while other fallbacks increase regret modestly; we therefore keep GDSF as the default in our experiments.
Under the fast-policy budget (\Cref{tab:overhead_budget}), the fallback shifts to the best-average fast policy (typically LHD).
For online selection on spliced traces, we apply minimal hysteresis (min-stay of one window) and a one-window lag
to avoid reacting to intra-window noise.

\paragraph{A simple ablation: \textsc{SCION-P90}.}
For comparison, we include a single-statistic selector using the prefix p90 size:
\[
  \pi(p90)=
  \begin{cases}
    \texttt{GDSF}, & p90 \ge \tau \\
    \texttt{S3-FIFO}, & p90 < \tau
  \end{cases}
\]
Here p90 is the 90th percentile of object size in the prefix and $\tau$ is a size threshold (bytes).
This mirrors the original mean-threshold intuition but is more robust under heavy tails.

\paragraph{Default hyperparameters.}
Unless otherwise stated, we fix $N{=}200$k, window size 200k, and use leave-one-trace-out tuning for $\tau$ in \textsc{SCION-P90}
over a small grid of thresholds.
In our evaluation, the expert set includes \texttt{GDSF}, \texttt{S3-FIFO}, \texttt{SIEVE}, \texttt{LHD}, \texttt{WTinyLFU-AV}, and \texttt{DynamicAdaptiveClimb}.
We also evaluate AdaptiveClimb as an additional baseline.
We train separate selectors per objective; \textsc{AUTO} targets OMR by default, while cost-weighted objectives use their own selectors.
\paragraph{Selector training details.}
The selector is trained offline to map fingerprints to the best expert under each objective, using held-out traces for validation.
We train a multinomial (softmax) linear classifier per objective on standardized features (z-scored per fold),
using batch gradient descent (800 steps, learning rate 0.1) with $L_2$ regularization ($10^{-3}$), and no per-trace reweighting.
Labels are the best-performing expert for each trace--size sample under the objective.
If multiple experts tie within $10^{-6}$, we mark the sample as \emph{flat} (degenerate) for analysis; label ties are broken deterministically by CSV order.
For leave-one-trace-out, each sample’s fallback policy is the best-average expert on the corresponding training fold.

\paragraph{Overhead.}
The fingerprint requires parsing only $N$ records once per trace/epoch. All per-request logic is inherited from the chosen expert.
This aligns with the minimal-overhead direction in recent learned caching work~\cite{mat,threeLcache}.

\section{Experimental Setup}
\label{sec:eval}
\paragraph{Simulator and implementation.}
We implement a trace-driven benchmark in C++ on top of \texttt{libCacheSim}~\cite{libcachesim}.
Experiments are CPU-only and generate CSV outputs, summary tables, and PDF figures.
Unless otherwise stated, \textsc{AUTO} makes a single selection using the first $N$ requests of each trace and keeps that expert for the full run;
only the spliced-trace experiments re-select at window boundaries with lag/min-stay.
\paragraph{Why CPU-only and trace-driven.}
``CPU-only'' here refers to the evaluation platform (no GPU/ML acceleration), not the provenance of traces.
Our traces come from CDN/object-cache workloads; trace-driven simulation provides reproducibility and isolates replacement effects,
and we complement it with a lightweight HTTP prototype to sanity-check end-to-end behavior.
\paragraph{Engineering scope.}
Beyond the core selector, we integrated AdaptiveClimb/DynamicAdaptiveClimb into \texttt{libCacheSim},
built a trace conversion pipeline to evaluate HR-Cache with matched windows,
and implemented end-to-end scripts for trace download, experiment sweeps, spliced-trace generation,
overhead measurement, and paper asset export.

\paragraph{Traces.}
We use 30 public object-cache traces from \texttt{cache\_dataset}~\cite{cache_dataset} (oracleGeneral format),
spanning Meta CDN workloads (reag/rprn/rnha), Wikipedia 2019, 10 Twitter clusters,
Alibaba/Tencent block caches, CloudPhysics traces, MSR traces, and Meta KV.
For broad coverage we cap each run at 5M requests and sweep cache sizes \{128, 256, 512, 1024\} MiB.
For four long traces (\texttt{meta\_reag}, \texttt{meta\_rprn}, \texttt{wiki\_2019t}, \texttt{twitter\_cluster10})
we additionally run 20M requests to check stability.
\Cref{tab:tracefeat} summarizes prefix statistics and cacheability/uniqueness diagnostics.
\paragraph{Data diagnostics and anomalies.}
We do not alter the public traces; instead, we surface diagnostics (prefix size percentiles, cacheable and unique ratios, and per-trace OMR ranges across policies)
and flag trace--size pairs where policies are indistinguishable (OMR range $\le 10^{-6}$) as degenerate.
These flagged pairs are excluded from win-count statistics but retained in summary averages for transparency.
Some Twitter clusters contain zero-size records in the public dataset, which leads to 0-valued percentiles in \Cref{tab:tracefeat};
we keep these entries and rely on the diagnostics to identify their impact. Because SCION uses percentile- and tail-based
features (not means), isolated zero-size records primarily affect the lower tail and have limited influence on p90-based fingerprints.
When percentiles are zero (a few Twitter clusters), we flag those trace--size pairs as degenerate for win-count statistics; the zeros
remain visible in the diagnostics and in the selector's input features.
Removing the three zero-size Twitter clusters (Twitter-10/13/35) shifts mean cacheable-only OMR by +0.0039 for \textsc{AUTO} and +0.0036 for GDSF,
and does not change the policy ordering (artifact).

\begin{table}[t]
  \centering
  \caption{Prefix trace diagnostics (size percentiles, cacheable and unique ratios). Sizes are in KiB/MiB; a few Twitter clusters include zero-size records in the public traces, which yields 0-sized percentiles and are flagged as degenerate for win-count statistics.}
  \label{tab:tracefeat}
  {\setlength{\tabcolsep}{3pt}\renewcommand{\arraystretch}{1.05}% Prefix size statistics and cacheable/unique ratios (N=prefix).
\scriptsize
\setlength{\tabcolsep}{3pt}
\renewcommand{\arraystretch}{0.9}
\resizebox{\linewidth}{!}{%
\begin{tabular}{@{}>{\raggedright\arraybackslash}p{0.30\linewidth}>{\raggedleft\arraybackslash}p{0.14\linewidth}>{\raggedleft\arraybackslash}p{0.14\linewidth}>{\raggedleft\arraybackslash}p{0.14\linewidth}>{\raggedleft\arraybackslash}p{0.14\linewidth}>{\raggedleft\arraybackslash}p{0.14\linewidth}@{}}
\toprule
Trace & p50 & p90 & Mean & Cacheable & Unique \\
\midrule
Alibaba-0 & 4.0{\scriptsize KiB} & 420.0{\scriptsize KiB} & 69.9{\scriptsize KiB} & 1.000 & 0.439 \\
Alibaba-1 & 4.0{\scriptsize KiB} & 8.0{\scriptsize KiB} & 6.2{\scriptsize KiB} & 1.000 & 0.159 \\
Alibaba-100 & 8.0{\scriptsize KiB} & 28.0{\scriptsize KiB} & 12.0{\scriptsize KiB} & 1.000 & 0.935 \\
Alibaba-103 & 12.0{\scriptsize KiB} & 72.0{\scriptsize KiB} & 29.5{\scriptsize KiB} & 1.000 & 0.778 \\
Alibaba-11 & 8.0{\scriptsize KiB} & 32.0{\scriptsize KiB} & 15.7{\scriptsize KiB} & 1.000 & 0.680 \\
Alibaba-12 & 8.0{\scriptsize KiB} & 504.0{\scriptsize KiB} & 89.4{\scriptsize KiB} & 1.000 & 0.984 \\
Meta-KV-1 & 0.1{\scriptsize KiB} & 0.4{\scriptsize KiB} & 0.5{\scriptsize KiB} & 1.000 & 0.180 \\
Meta-reag & 26.8{\scriptsize KiB} & 8.2{\scriptsize MiB} & 23.0{\scriptsize MiB} & 0.989 & 0.428 \\
Meta-rnha & 64.6{\scriptsize KiB} & 20.1{\scriptsize MiB} & 41.0{\scriptsize MiB} & 0.985 & 0.688 \\
Meta-rprn & 55.7{\scriptsize KiB} & 10.0{\scriptsize MiB} & 32.3{\scriptsize MiB} & 0.992 & 0.575 \\
MSR-prn-0 & 4.0{\scriptsize KiB} & 64.0{\scriptsize KiB} & 14.1{\scriptsize KiB} & 1.000 & 0.395 \\
MSR-proj-0 & 4.0{\scriptsize KiB} & 64.0{\scriptsize KiB} & 14.0{\scriptsize KiB} & 1.000 & 0.268 \\
MSR-prxy-0 & 1.0{\scriptsize KiB} & 4.0{\scriptsize KiB} & 2.8{\scriptsize KiB} & 1.000 & 0.054 \\
Tencent-10000 & 4.0{\scriptsize KiB} & 12.0{\scriptsize KiB} & 8.7{\scriptsize KiB} & 1.000 & 0.290 \\
Tencent-10001 & 4.0{\scriptsize KiB} & 112.0{\scriptsize KiB} & 22.5{\scriptsize KiB} & 1.000 & 0.745 \\
Tencent-10100 & 4.0{\scriptsize KiB} & 32.0{\scriptsize KiB} & 20.7{\scriptsize KiB} & 1.000 & 0.626 \\
Tencent-10348 & 4.0{\scriptsize KiB} & 40.0{\scriptsize KiB} & 18.8{\scriptsize KiB} & 1.000 & 0.384 \\
Tencent-10512 & 4.0{\scriptsize KiB} & 32.0{\scriptsize KiB} & 18.4{\scriptsize KiB} & 1.000 & 0.629 \\
Tencent-10533 & 4.0{\scriptsize KiB} & 48.0{\scriptsize KiB} & 17.7{\scriptsize KiB} & 1.000 & 0.085 \\
Twitter-10 & 0.0{\scriptsize KiB} & 0.0{\scriptsize KiB} & 0.0{\scriptsize KiB} & 1.000 & 0.500 \\
Twitter-13 & 0.0{\scriptsize KiB} & 3.8{\scriptsize KiB} & 2.3{\scriptsize KiB} & 1.000 & 0.629 \\
Twitter-20 & 0.1{\scriptsize KiB} & 0.1{\scriptsize KiB} & 0.1{\scriptsize KiB} & 1.000 & 0.139 \\
Twitter-26 & 0.1{\scriptsize KiB} & 0.2{\scriptsize KiB} & 0.2{\scriptsize KiB} & 1.000 & 0.077 \\
Twitter-3 & 0.1{\scriptsize KiB} & 0.3{\scriptsize KiB} & 0.2{\scriptsize KiB} & 1.000 & 0.075 \\
Twitter-35 & 0.0{\scriptsize KiB} & 5.0{\scriptsize KiB} & 1.3{\scriptsize KiB} & 1.000 & 0.077 \\
CloudPhysics-w01 & 3.5{\scriptsize KiB} & 6.5{\scriptsize KiB} & 16.1{\scriptsize KiB} & 1.000 & 0.870 \\
CloudPhysics-w02 & 16.0{\scriptsize KiB} & 128.0{\scriptsize KiB} & 47.8{\scriptsize KiB} & 1.000 & 0.335 \\
CloudPhysics-w03 & 32.0{\scriptsize KiB} & 256.0{\scriptsize KiB} & 71.6{\scriptsize KiB} & 1.000 & 0.608 \\
CloudPhysics-w04 & 32.0{\scriptsize KiB} & 256.0{\scriptsize KiB} & 79.9{\scriptsize KiB} & 1.000 & 0.493 \\
Wiki-2019t & 22.0{\scriptsize KiB} & 70.4{\scriptsize KiB} & 33.3{\scriptsize KiB} & 1.000 & 0.707 \\
\bottomrule
\end{tabular}
}
\normalsize
}
\end{table}

\paragraph{Metrics and ``cacheable-only'' reporting.}
Traces can contain objects larger than the cache capacity; those requests are effectively uncacheable and can distort conclusions.
We report cacheable-only object miss ratio (OMR) and cacheable-only byte miss ratio (BMR),
computed over requests/bytes restricted to objects whose size fits within the evaluated cache capacity.
We also report all-request OMR/BMR (including uncacheable objects) for operational interpretation.
OMR is a proxy for hit ratio and user-facing latency, while BMR captures backend bandwidth and cost;
in object caches, large objects can dominate bytes even when OMR is low, so we report both.
We compute bootstrap confidence intervals (CIs) from windowed miss ratios, and report trace integrity
statistics (size percentiles, cacheable ratio, unique ratio) to identify degenerate trace--size pairs.
Because cacheable-only and all-request objectives can diverge, we include concrete tradeoff examples in the reproducibility artifact.

\paragraph{Compared policies.}
We report \textsc{AUTO} (selector over \{GDSF, S3-FIFO, SIEVE, LHD, W-TinyLFU-AV, DynamicAdaptiveClimb~\cite{dynamicadaptiveclimb}\}) and its experts,
and compare against additional baselines including AdaptiveClimb~\cite{dynamicadaptiveclimb}, W-TinyLFU, AdaptSize-style admission
(admit with probability $\exp(-\texttt{size}/C)$; we tune $C$ with a small grid per trace),
and classic policies such as ARC, TwoQ, LIRS, LeCaR, and Cacheus (supported by our benchmark harness).
We also evaluate HR-Cache (hazard-rate cache)~\cite{hrcache} using its open-source simulator on a 10-trace subset and two cache sizes
(256/1024\,MiB), converting traces to its input format (32-bit ids/sizes) and matching our 200k-window evaluation cadence.

\paragraph{System prototype.}
To validate feasibility beyond trace-driven simulation, we implement a lightweight end-to-end prototype:
an in-memory cache daemon (this work) serves HTTP GET requests and fetches misses from a local origin server, with optional TTLs but no revalidation logic.
The daemon uses the same policy logic as the simulator (GDSF, S3-FIFO, SIEVE, and \textsc{AUTO}),
and exposes hit ratio and origin byte counters.
We replay a smaller prefix of each trace to keep the run time reasonable and report latency, throughput, and resource usage.
We optionally inject origin-side delay to emulate network jitter.
For the prototype, \textsc{AUTO} uses the p90-threshold selector (\textsc{SCION-P90}) to avoid offline training dependencies.
Unless otherwise stated, the prototype uses a 256\,MiB cache, replays 20k requests per trace, and issues concurrent GETs with client concurrency 128.
This prototype omits revalidation, invalidations, and distributed cache behavior; it is intended as a feasibility check rather than a production model.

\section{Results}
\label{sec:results}
\subsection{Aggregate performance}
\Cref{tab:avg} summarizes average cacheable-only miss ratios.
Across workloads, \textsc{AUTO} is competitive with the strongest experts on OMR while reducing the risk of regime mismatch;
its average OMR remains close to the best-performing expert, but avoids large mismatches when the workload falls into a different regime.
GDSF remains the strongest average-OMR baseline on this suite.
We compute 95\% bootstrap confidence intervals (CIs) over trace--size pairs for key policies and report them in \Cref{tab:avg},
highlighting that the observed differences are often small but consistent.
All-request averages show the same ordering: AUTO OMR/BMR 0.439/0.575 versus GDSF 0.433/0.577 and SIEVE 0.479/0.596,
and \Cref{tab:tradeoffsum} summarizes cacheable-only OMR vs all-request BMR tradeoffs.
Against GDSF, we do not observe the specific conflict ``better cacheable-only OMR but worse all-request BMR'';
operationally, the usual tradeoff is instead slightly higher OMR for slightly lower BMR.
\paragraph{Fast-policy budget as a primary operating point.}
For CPU-constrained deployments we treat the fast-policy budget as a primary operating point (not just a side analysis).
The fast set excludes WTinyLFU-AV because its single-threaded throughput is $<1\%$ of SIEVE in our implementation, so fast-policy conclusions are not driven by that outlier.
Production TinyLFU variants (e.g., Caffeine/Ristretto) and cost-aware GDS approximations are not yet available in our simulator; they are natural fast-policy additions for future SCION deployments.
\Cref{tab:overhead_budget} reports \textsc{AUTO-fast} against the best fixed fast policy and the fast-set oracle;
\textsc{AUTO-fast} trades a small OMR increase for a larger BMR reduction, yielding a lower $\alpha{=}0.5$ cost than the fixed fast baseline.
\paragraph{Why OMR improves.}
On large-object regimes, size-aware experts reduce object misses among cacheable objects, while FIFO-family designs can be misled by object-size heterogeneity.
On small-object regimes, S3-FIFO is already near-optimal among our compared baselines; \textsc{SCION} simply selects it, avoiding degradation.
\paragraph{Case studies (with CIs).}
Example (CloudPhysics-w02, 1\,GiB): \textsc{AUTO} selects GDSF, achieving cacheable-only OMR 0.438 [0.428, 0.454] versus SIEVE 0.976 [0.971, 0.981],
and BMR 0.888 vs 0.995 (10.7 percentage points lower).
Example (Alibaba-103, 256\,MiB): \textsc{AUTO} selects WTinyLFU-AV, achieving cacheable-only OMR 0.768 [0.758, 0.784] versus GDSF 0.823 [0.814, 0.835],
and BMR 0.872 vs 0.914.
More per-trace examples are in the reproducibility artifact.

\begin{table}[htbp]
  \centering
  \caption{Representative per-trace examples with OMR CIs (AUTO vs baseline).}
  \label{tab:trace_examples}
  {% Representative per-trace examples with OMR CIs (AUTO vs baseline).
\small
\setlength{\tabcolsep}{3pt}\renewcommand{\arraystretch}{1.05}
\resizebox{\linewidth}{!}{%
\begin{tabular}{@{}llllll@{}}
\toprule
Trace & Cache & AUTO policy & AUTO OMR [CI] & Baseline & Baseline OMR [CI] \\
\midrule
CloudPhysics-w02 & 1GiB & GDSF & 0.438 [0.428,0.454] & SIEVE & 0.976 [0.971,0.981] \\
Alibaba-103 & 256MiB & WTinyLFU-AV & 0.768 [0.758,0.784] & GDSF & 0.823 [0.814,0.835] \\
CloudPhysics-w02 & 128MiB & GDSF & 0.979 [0.975,0.984] & WTinyLFU-AV & 0.734 [0.714,0.758] \\
\bottomrule
\end{tabular}%
}
}
\end{table}

\begin{table}[htbp]
  \centering
  \caption{Average cacheable-only miss ratios over all evaluated workloads with 95\% bootstrap CIs. Lower is better.}
  \label{tab:avg}
  {\setlength{\tabcolsep}{3pt}\renewcommand{\arraystretch}{1.05}% Average cacheable-only miss ratios with 95\% bootstrap CIs.
\resizebox{\linewidth}{!}{%
\begin{tabular}{@{}>{\raggedright\arraybackslash}p{0.32\linewidth}>{\raggedleft\arraybackslash}p{0.34\linewidth}>{\raggedleft\arraybackslash}p{0.34\linewidth}@{}}
\toprule
Policy & Avg OMR (95\% CI) & Avg BMR (95\% CI) \\
\midrule
AUTO & 0.438240 [0.388795,0.490928] & 0.528066 [0.469773,0.586067] \\
SIEVE & 0.479133 [0.423313,0.532635] & 0.541692 [0.484306,0.600559] \\
S3-FIFO & 0.468178 [0.413922,0.521532] & 0.566012 [0.511392,0.624286] \\
GDSF & 0.432585 [0.381961,0.484499] & 0.530435 [0.471803,0.588821] \\
DynamicAdaptiveClimb & 0.470352 [0.415332,0.524424] & 0.532683 [0.473753,0.592131] \\
\bottomrule
\end{tabular}
}
}
\end{table}

\begin{figure*}[t]
  \centering
  \includegraphics[width=0.97\textwidth]{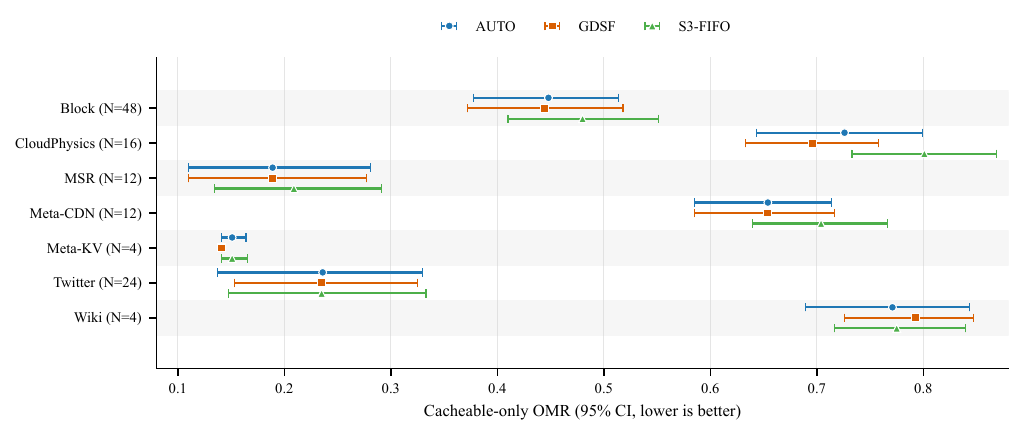}
  \caption{Family-level cacheable-only OMR with 95\% CIs. The regime split is stable across families: CloudPhysics favors GDSF, Wiki favors AUTO, and FIFO-family policies trail on the large-object families.}
  \label{fig:family_summary_plot}
\end{figure*}

\paragraph{Per-family summary.}
\Cref{tab:family_summary} aggregates cacheable-only OMR by dataset family (averaged over trace--size pairs with bootstrap CIs),
surfacing consistent regime differences across families.
We report \textsc{AUTO}, a size-aware baseline (GDSF), and a FIFO-family baseline (S3-FIFO); full per-trace results remain in the artifact.
Figure~\ref{fig:family_summary_plot} provides the same comparison visually and makes the regime split across families easier to scan.

\begin{table}[htbp]
  \centering
  \caption{Family-level cacheable-only OMR summary (95\% CIs).}
  \label{tab:family_summary}
  {\setlength{\tabcolsep}{3pt}\renewcommand{\arraystretch}{1.02}% Family-level cacheable-only OMR summary with 95% CIs.
\resizebox{\linewidth}{!}{%
\begin{tabular}{@{}>{\raggedright\arraybackslash}p{0.20\linewidth}>{\raggedleft\arraybackslash}p{0.08\linewidth}>{\raggedleft\arraybackslash}p{0.22\linewidth}>{\raggedleft\arraybackslash}p{0.22\linewidth}>{\raggedleft\arraybackslash}p{0.22\linewidth}@{}}
\toprule
Family & $N$ & AUTO & GDSF & S3-FIFO \\
\midrule
Block & 48 & 0.448 [0.378,0.514] & 0.444 [0.372,0.518] & 0.480 [0.410,0.551] \\
CloudPhysics & 16 & 0.726 [0.643,0.799] & 0.696 [0.633,0.758] & 0.801 [0.733,0.869] \\
MSR & 12 & 0.189 [0.110,0.281] & 0.189 [0.110,0.277] & 0.209 [0.134,0.291] \\
Meta-CDN & 12 & 0.654 [0.585,0.714] & 0.654 [0.585,0.717] & 0.704 [0.640,0.766] \\
Meta-KV & 4 & 0.151 [0.141,0.164] & 0.141 [0.141,0.142] & 0.151 [0.141,0.165] \\
Twitter & 24 & 0.236 [0.137,0.330] & 0.235 [0.153,0.325] & 0.235 [0.148,0.333] \\
Wiki & 4 & 0.771 [0.689,0.843] & 0.793 [0.726,0.847] & 0.775 [0.717,0.840] \\
\bottomrule
\end{tabular}
}
}
\end{table}

\paragraph{AUTO vs strong baselines.}
\Cref{tab:auto_vs_baselines} summarizes AUTO--baseline OMR deltas (negative favors AUTO).
Averaged over all trace--size pairs, AUTO improves S3-FIFO on 65.8\% of pairs (avg $\Delta{=}{-}0.0299$) and improves GDSF on 13.3\%
of pairs (avg $\Delta{=}+0.0057$), illustrating the regime-mismatch tradeoff with a very strong size-aware baseline.
Against GDSF, 72/120 pairs are ties. Family-wise, AUTO is best on Wiki ($\Delta$OMR $-0.0215$, $\Delta$BMR $-0.0468$),
roughly tied on Meta/Twitter/MSR, and most clearly loses on CloudPhysics ($\Delta$OMR $+0.0299$),
where a single expert dominates across sizes; block traces sit in between with a small average OMR loss ($+0.0044$) but average BMR gain ($-0.0062$).
The largest overall AUTO gaps occur on CloudPhysics w02/w01 (avg gaps 0.133/0.130; max gaps 0.246/0.153), so these are natural ``pin GDSF'' regimes.

\begin{table}[htbp]
  \centering
\caption{Key aggregate gains vs strong baselines on cacheable-only OMR (delta = AUTO -- baseline).}
  \label{tab:auto_vs_baselines}
  {\setlength{\tabcolsep}{3pt}\renewcommand{\arraystretch}{1.05}% AUTO vs baselines summary (cacheable-only OMR; delta = AUTO - baseline)
\small
\setlength{\tabcolsep}{5pt}\renewcommand{\arraystretch}{1.05}
\begin{tabular}{lrrrr}
\toprule
Baseline & Improve\% & Avg $\Delta$ & Best $\Delta$ & Worst $\Delta$ \\
\midrule
GDSF & 13.3 & 0.0057 & -0.2030 & 0.1402 \\
S3-FIFO & 65.8 & -0.0299 & -0.5426 & 0.0841 \\
\bottomrule
\end{tabular}
\normalsize
}
\end{table}
CI computation bootstraps the per trace--size averages (500 resamples) rather than individual requests/windows,
preserving within-trace dependence by treating each trace--size pair as the resampling unit.

\paragraph{SCION-P90 ablation.}
\textsc{SCION-P90} (artifact table) is competitive but trails \textsc{AUTO} on average OMR/BMR across all trace--size pairs.
AUTO reduces cacheable-only OMR by $\approx$0.020 and BMR by $\approx$0.006 relative to \textsc{SCION-P90},
indicating that the multi-feature selector adds value beyond a single threshold.
\textsc{SCION-P90} matches \textsc{AUTO} on 21.7\% of pairs overall; because \textsc{SCION-P90} can only choose between GDSF and S3-FIFO,
it matches \textsc{AUTO} on 48.1\% of the pairs where \textsc{AUTO} selects one of those two experts (artifact).
\textsc{AUTO} helps most when p90 is near $\tau$ (artifact), where cacheability/uniqueness features provide additional separation (\cref{tab:selector_ablation}).

\paragraph{HR-Cache subset.}
\Cref{tab:hrcache} compares HR-Cache~\cite{hrcache} against our baselines on the overlapping subset
(10 traces, 256/1024\,MiB). HR-Cache achieves competitive BMR on this subset; the BMR-aligned selector AUTO($\alpha{=}0.2$) closes much of the gap,
while \textsc{AUTO} and our strongest non-ML baselines
remain strong on OMR.
HR-Cache expects 32-bit signed IDs/sizes; our conversion maps IDs via modulo and clamps sizes to that range, which can introduce collisions.
Accordingly, we treat HR-Cache results as a cross-check rather than a definitive head-to-head ranking.
The subset reflects the extra conversion/build steps required by the HR-Cache simulator; we provide scripts to extend it further.

\paragraph{Classic baselines.}
\Cref{tab:classic} reports ARC and TwoQ on a 6-trace subset (smallest HR-Cache traces, 256/1024\,MiB),
using the first 20k requests per trace to keep runtime manageable for these heavier policies.
LIRS is substantially slower in our simulator (minutes per 20k requests on mid-sized traces), so we report it on the two smallest traces
using 5k-request prefixes; the table includes per-policy sample counts (N).
These classical policies are effectively tied on this small subset; because the prefixes are much shorter than the main suite,
these values are not directly comparable to full-suite averages and are intended as qualitative anchors.
We do not see evidence that they dominate
the best size-aware experts on cacheable-only OMR.

\begin{table}[htbp]
  \centering
  \caption{Classic baselines on a small HR-Cache subset (cacheable-only averages; 20k-request prefixes for ARC/TwoQ, 5k for LIRS; N shows samples).}
  \label{tab:classic}
  {\setlength{\tabcolsep}{3pt}\renewcommand{\arraystretch}{1.05}% Classic baselines on HR-Cache subset (cacheable-only averages)
\resizebox{\linewidth}{!}{%
\begin{tabular}{@{}>{\raggedright\arraybackslash}p{0.36\linewidth}>{\raggedleft\arraybackslash}p{0.22\linewidth}>{\raggedleft\arraybackslash}p{0.22\linewidth}>{\raggedleft\arraybackslash}p{0.12\linewidth}@{}}
\toprule
Policy & Avg OMR $\downarrow$ & Avg BMR $\downarrow$ & N \\
\midrule
ARC & 0.382226 & 0.448056 & 12 \\
TwoQ & 0.379894 & 0.472687 & 12 \\
LIRS & 0.422354 & 0.529858 & 4 \\
\bottomrule
\end{tabular}
}
}
\end{table}

\paragraph{Learned baselines subset.}
To position against learned baselines, we evaluate LeCaR, Cacheus, and LRB-BMR (LightGBM-based learned replacement optimizing BMR)
on the 10-trace subset using 200k requests and four cache sizes (128/256/512/1024\,MiB). \Cref{tab:learned_subset} shows that
LeCaR attains the best average OMR on the successful 32-pair subset (0.504), while the BMR-oriented selector AUTO($\alpha{=}0.2$)
achieves the lowest average BMR in our harness (0.466 vs 0.472 for LRB-BMR) with lower average OMR than LRB-BMR (0.520 vs 0.530).
Default AUTO remains close to GDSF on OMR (0.513 vs 0.510), illustrating that SCION's value is exposing explicit operating points rather than dominating every specialized learned baseline on its home objective.
This limited subset is meant as a qualitative anchor rather than a full-suite comparison; full results and scripts are in the artifact.

\begin{table}[htbp]
  \centering
  \caption{Learned baselines on the 10-trace subset (cacheable-only averages; 200k requests; 128/256/512/1024\,MiB).}
  \label{tab:learned_subset}
  {\setlength{\tabcolsep}{3pt}\renewcommand{\arraystretch}{1.05}% Learned baselines on HR subset.
\resizebox{\linewidth}{!}{%
\begin{tabular}{@{}>{\raggedright\arraybackslash}p{0.36\linewidth}>{\raggedleft\arraybackslash}p{0.22\linewidth}>{\raggedleft\arraybackslash}p{0.22\linewidth}>{\raggedleft\arraybackslash}p{0.12\linewidth}@{}}
\toprule
Policy & Avg OMR $\downarrow$ & Avg BMR $\downarrow$ & N \\
\midrule
AUTO & 0.513082 & 0.496279 & 40 \\
AUTO ($\alpha=0.2$) & 0.520135 & 0.465623 & 40 \\
GDSF & 0.510144 & 0.507296 & 40 \\
LeCaR & 0.504318 & 0.507363 & 32 \\
Cacheus & 0.526992 & 0.482302 & 40 \\
LRB-\allowbreak{}BMR & 0.529727 & 0.471793 & 40 \\
\bottomrule
\end{tabular}
}
}
\end{table}

\paragraph{Oracle lookahead bounds.}
To contextualize headroom, we also report oracle lookahead policies (Belady and BeladySize) on a 6-trace subset
using 20k-request prefixes (oracleGeneral traces include next-access timestamps).
\Cref{tab:opt_bounds} provides cacheable-only averages; these act as lightweight object- and size-aware lower bounds.
On this subset, BeladySize reaches 0.372 OMR / 0.447 BMR, versus 0.375 / 0.466 for the strongest online expert (GDSF),
so the residual gap is modest for OMR but clearer for byte-oriented objectives.
Belady lowers OMR further (0.346) while worsening BMR relative to BeladySize, reinforcing that object-optimal and byte-aware offline comparators differ under variable sizes.
Tighter flow-based bounds such as FOO/PFOO are therefore complementary~\cite{foo_pfoo}, but require a different harness, so we leave them for future work.

\begin{table}[htbp]
  \centering
  \caption{Oracle lookahead bounds on a small subset (cacheable-only averages; 20k-request prefixes).}
  \label{tab:opt_bounds}
  {\setlength{\tabcolsep}{3pt}\renewcommand{\arraystretch}{1.05}% Oracle lookahead bounds on a small subset (cacheable-only averages)
\resizebox{\linewidth}{!}{%
\begin{tabular}{@{}>{\raggedright\arraybackslash}p{0.44\linewidth}>{\raggedleft\arraybackslash}p{0.23\linewidth}>{\raggedleft\arraybackslash}p{0.23\linewidth}@{}}
\toprule
Policy & Avg OMR $\downarrow$ & Avg BMR $\downarrow$ \\
\midrule
Belady & 0.377158 & 0.445614 \\
BeladySize & 0.372115 & 0.447127 \\
\bottomrule
\end{tabular}
}
}
\end{table}

\begin{table}[htbp]
  \centering
  \caption{Average cacheable-only miss ratios on the HR-Cache subset.}
  \label{tab:hrcache}
  {\setlength{\tabcolsep}{3pt}\renewcommand{\arraystretch}{1.05}% Average cacheable-only miss ratios on the HR-Cache subset (10 traces, 2 cache sizes).
\resizebox{\linewidth}{!}{%
\begin{tabular}{@{}>{\raggedright\arraybackslash}p{0.44\linewidth}>{\raggedleft\arraybackslash}p{0.23\linewidth}>{\raggedleft\arraybackslash}p{0.23\linewidth}@{}}
\toprule
Policy & Avg OMR $\downarrow$ & Avg BMR $\downarrow$ \\
\midrule
HR-Cache & 0.504266 & 0.457858 \\
AUTO ($\alpha=0.2$) & 0.490309 & 0.452126 \\
AUTO & 0.483891 & 0.482311 \\
SIEVE & 0.508924 & 0.450670 \\
S3-FIFO & 0.502109 & 0.558957 \\
GDSF & 0.475218 & 0.487148 \\
DynamicAdaptiveClimb & 0.499772 & 0.435142 \\
\bottomrule
\end{tabular}
}
}
\end{table}

\paragraph{Selector accuracy.}
\Cref{tab:selector} reports leave-one-trace-out selector accuracy and regret distributions.
We define \emph{regret} as the per-sample objective gap between the selected policy and the best expert
(e.g., $\mathrm{OMR}_{\text{chosen}} - \min_e \mathrm{OMR}_e$), and mark a sample as \emph{nonflat} if
the best--worst gap exceeds $10^{-6}$.
The selector is accurate on non-degenerate trace--size pairs and exhibits low average regret.
Per-trace avg regret has median 0.0079 and P90 0.0675; per-trace p90-regret median 0.0126 (P90 0.1003) (artifact),
indicating most traces are low-regret with a small tail of difficult cases.
Low-confidence triggers are concentrated in a few traces (artifact); \Cref{tab:selector} reports low-confidence fractions by objective,
with 16.3\% of non-flat OMR pairs falling back in the final \textsc{AUTO} setting (leave-one-trace-out).
Top-1 calibration on non-flat OMR samples yields ECE 0.10 and Brier 0.21, indicating mild overconfidence; temperature scaling does not improve regret (artifact).
A threshold sweep in the 6-expert sensitivity study shows the coverage--regret tradeoff and motivates $\theta{=}0.40$ (artifact).

\begin{table}[htbp]
  \centering
  \caption{Selector accuracy and regret (lower regret is better).}
  \label{tab:selector}
  {\setlength{\tabcolsep}{3pt}\renewcommand{\arraystretch}{1.05}% Selector accuracy and regret (lower regret is better).
\resizebox{\linewidth}{!}{%
\begin{tabular}{@{}>{\raggedright\arraybackslash}p{0.22\linewidth}>{\raggedright\arraybackslash}p{0.12\linewidth}>{\raggedleft\arraybackslash}p{0.08\linewidth}>{\raggedleft\arraybackslash}p{0.14\linewidth}>{\raggedleft\arraybackslash}p{0.14\linewidth}>{\raggedleft\arraybackslash}p{0.14\linewidth}>{\raggedleft\arraybackslash}p{0.12\linewidth}@{}}
\toprule
Objective & Subset & Acc & Avg regret & P95 regret & Worst regret & Low-conf\% \\
\midrule
cost:0.\allowbreak{}2 & all & 0.433 & 0.024102 & 0.113755 & 0.182123 & 42.5 \\
cost:0.\allowbreak{}2 & nonflat & 0.327 & 0.029513 & 0.116286 & 0.182123 & 51.0 \\
cost:0.\allowbreak{}5 & all & 0.508 & 0.020515 & 0.106438 & 0.189169 & 42.5 \\
cost:0.\allowbreak{}5 & nonflat & 0.418 & 0.025121 & 0.109354 & 0.189169 & 51.0 \\
cost:0.\allowbreak{}8 & all & 0.558 & 0.015684 & 0.096900 & 0.203912 & 16.7 \\
cost:0.\allowbreak{}8 & nonflat & 0.500 & 0.019205 & 0.106430 & 0.203912 & 17.3 \\
omr & all & 0.525 & 0.019668 & 0.110751 & 0.245795 & 14.2 \\
omr & nonflat & 0.459 & 0.024084 & 0.119545 & 0.245795 & 16.3 \\
\bottomrule
\end{tabular}
}
}
\end{table}

\paragraph{Model sensitivity.}
We compare the linear softmax selector (standardized features) to an unstandardized variant and a tiny one-hidden-layer MLP.
Model sensitivity results are reported in the reproducibility artifact; standardization matters for linear models,
and a tiny MLP slightly improves average regret while increasing tail regret, so we keep the linear model for stability.

\paragraph{Selector ablations and fallback behavior.}
\Cref{tab:selector_ablation} drops one feature at a time; cacheability, uniqueness, and cache size contribute the largest regret increases,
while p50/p90/mean/tail-ratio are comparatively redundant on this trace suite.
A sweep over confidence thresholds (artifact) selects $\theta{=}0.40$; average regret varies by $<0.005$ across $\theta\in[0.3,0.6]$.
Temperature scaling does not change coverage or regret at $\theta{=}0.40$, so we report uncalibrated scores.
Perturbing cacheable/unique ratios by 2--30\% relative noise increases average regret by only 0.002--0.003 (artifact),
consistent with sketching errors from HLL/t-digest-style estimators and showing headroom beyond typical 2--10\% HLL-scale noise.
We also report per-objective selector weights in the artifact; the largest coefficients align with cacheability/uniqueness and size percentiles,
consistent with the regime split observed in the ablations.
\Cref{tab:fallback_sensitivity} summarizes objective-specific fallback sensitivity; richer uncertainty triggers are future work.

\begin{table}[htbp]
  \centering
\caption{Fallback sensitivity across objectives (non-flat samples; DynAdaptiveClimb = DynamicAdaptiveClimb).}
  \label{tab:fallback_sensitivity}
  {\setlength{\tabcolsep}{3pt}\renewcommand{\arraystretch}{1.02}% Fallback sensitivity across objectives (non-flat samples).
\resizebox{\linewidth}{!}{%
\begin{tabular}{@{}>{\raggedright\arraybackslash}p{0.30\linewidth}>{\raggedleft\arraybackslash}p{0.17\linewidth}>{\raggedleft\arraybackslash}p{0.17\linewidth}>{\raggedleft\arraybackslash}p{0.17\linewidth}>{\raggedleft\arraybackslash}p{0.17\linewidth}@{}}
\toprule
Fallback & OMR avg & OMR P95 & $\alpha{=}0.2$ avg & $\alpha{=}0.2$ P95 \\
\midrule
GDSF & 0.0241 & 0.1195 & 0.0295 & 0.1163 \\
DynAdaptiveClimb & 0.0325 & 0.1218 & 0.0476 & 0.1627 \\
LHD & 0.0248 & 0.1195 & 0.0441 & 0.1638 \\
S3-FIFO & 0.0300 & 0.1277 & 0.0396 & 0.1401 \\
SIEVE & 0.0321 & 0.1277 & 0.0529 & 0.1884 \\
\bottomrule
\end{tabular}
}
}
\end{table}

\paragraph{Expert-set ablation.}
\Cref{tab:expert_ablation} drops one expert at a time and retrains the selector.
We report the selector regret within the reduced set, the \emph{oracle gap} (loss from removing an expert, relative to the full-set best),
and the combined total gap.
Dropping GDSF increases the oracle gap most, while LHD and WTinyLFU-AV also contribute non-trivial headroom in their regimes.
Dropping WTinyLFU-AV changes the total gap only slightly (0.0152 vs 0.0158), indicating that conclusions are not driven by its slow implementation.
Enumerating all 3- and 4-expert subsets (artifact) shows the best 4-expert subset has total gap 0.0130 and the best 3-expert subset
has 0.0148, indicating modest loss when operators restrict implementations.

\paragraph{Error analysis.}
We bucket non-flat samples by feature quantiles (artifact).
Misclassification rates are higher in small-object regimes (low p90) and low-unique-ratio regimes,
while mid p90 bins exhibit lower regret, suggesting that additional reuse/burstiness features could further stabilize selection in small-object workloads.
A lightweight reuse/burstiness sketch (e.g., a small count-min or recent-reuse histogram) is a plausible low-overhead extension.

\begin{table}[htbp]
  \centering
  \caption{Drop-one-feature ablation (non-flat samples).}
  \label{tab:selector_ablation}
  {\setlength{\tabcolsep}{3pt}\renewcommand{\arraystretch}{1.05}% Drop-one-feature ablation (non-flat samples).
\resizebox{\linewidth}{!}{%
\begin{tabular}{@{}>{\raggedright\arraybackslash}p{0.45\linewidth}>{\raggedleft\arraybackslash}p{0.25\linewidth}>{\raggedleft\arraybackslash}p{0.20\linewidth}@{}}
\toprule
Setting & Avg regret & Misclass rate \\
\midrule
all & 0.015819 & 0.367 \\
drop log p50 & 0.016706 & 0.367 \\
drop log p90 & 0.015819 & 0.367 \\
drop log mean & 0.015967 & 0.357 \\
drop log tail ratio & 0.016721 & 0.367 \\
drop cacheable ratio & 0.018268 & 0.367 \\
drop unique ratio & 0.018015 & 0.398 \\
drop log cache size & 0.018076 & 0.357 \\
\bottomrule
\end{tabular}
}
}
\end{table}

\begin{table}[htbp]
  \centering
  \caption{Expert-set ablation (cacheable-only OMR).}
  \label{tab:expert_ablation}
  {\setlength{\tabcolsep}{3pt}\renewcommand{\arraystretch}{1.05}% Expert-set ablation (cacheable-only OMR).
\resizebox{\linewidth}{!}{%
\begin{tabular}{@{}>{\raggedright\arraybackslash}p{0.36\linewidth}>{\raggedleft\arraybackslash}p{0.20\linewidth}>{\raggedleft\arraybackslash}p{0.20\linewidth}>{\raggedleft\arraybackslash}p{0.20\linewidth}@{}}
\toprule
Setting & Avg regret & Oracle gap & Total gap \\
\midrule
all & 0.0158 & 0.0000 & 0.0158 \\
drop GDSF & 0.0152 & 0.0098 & 0.0249 \\
drop DynamicAdaptiveClimb & 0.0178 & 0.0000 & 0.0178 \\
drop LHD & 0.0150 & 0.0021 & 0.0171 \\
drop WTinyLFU-AV & 0.0086 & 0.0066 & 0.0152 \\
drop S3-FIFO & 0.0145 & 0.0000 & 0.0145 \\
drop SIEVE & 0.0145 & 0.0000 & 0.0145 \\
\bottomrule
\end{tabular}
}
}
\end{table}

\paragraph{Fallback frequency in practice.}
Using the actual \textsc{AUTO} selector (leave-one-trace-out, $\theta{=}0.40$), low-confidence cases account for 16.3\% of non-flat
trace--size pairs, and all of those fall back to GDSF; nevertheless, \textsc{AUTO} selects a non-GDSF expert on 49.0\% of non-flat pairs.
Low-confidence fractions vary by family (MSR 0.36, block 0.21, CloudPhysics 0.12, and 0.0 for Meta-CDN/Twitter/Wiki);
when fallback triggers, 68.8\% of cases are within 0.01 regret and 87.5\% within 0.02 (artifact).

\paragraph{OMR/BMR tradeoff.}
\Cref{tab:cost} shows average cost for $\alpha \in \{0.2,0.5,0.8\}$ in $\alpha\cdot\mathrm{OMR} + (1-\alpha)\cdot\mathrm{BMR}$,
demonstrating that \textsc{SCION} can explicitly target operator-weighted objectives.
For example, AUTO($\alpha{=}0.5$) yields a slightly lower average cost than fixed GDSF (0.48126 vs 0.48151).
We also summarize how often cacheable-only OMR improvements coincide with worse all-request BMR in \Cref{tab:tradeoffsum}.
Case study (meta\_rnha, 1\,GiB): AUTO($\alpha{=}0.2$) selects DynamicAdaptiveClimb (OMR 0.784, all-request BMR 0.841),
while AUTO($\alpha{=}0.8$) selects GDSF (OMR 0.737, all-request BMR 0.886), illustrating the operator-controlled tradeoff.
Cross-objective robustness is reasonable: the $\alpha{=}0.5$ selector is within 0.0007 of the $\alpha{=}0.2$-tuned cost
and within 0.0035 of the $\alpha{=}0.8$-tuned cost, suggesting modest sensitivity to operator weight drift.
This supports a simple deployment strategy: pretrain a small set of selectors at a few $\alpha$ values and switch among them
as preferences drift; the $\alpha{=}0.2$ and $\alpha{=}0.8$ selectors choose different experts on 43.3\% of trace--size pairs (artifact).
Per-family and worst-case deltas for objective mismatch are included in the reproducibility artifact.

\begin{table}[htbp]
  \centering
  \caption{Avg. cost tradeoff for $\alpha$-weighted objective. Rows denote selector trained with $\alpha$; columns evaluate cost using $\alpha$.}
  \label{tab:cost}
  {\setlength{\tabcolsep}{3pt}\renewcommand{\arraystretch}{1.05}% Average cost $\alpha\cdot\mathrm{OMR} + (1-\alpha)\cdot\mathrm{BMR}$.
\resizebox{\linewidth}{!}{%
\begin{tabular}{@{}>{\raggedright\arraybackslash}p{0.36\linewidth}>{\raggedleft\arraybackslash}p{0.16\linewidth}>{\raggedleft\arraybackslash}p{0.16\linewidth}>{\raggedleft\arraybackslash}p{0.16\linewidth}@{}}
\toprule
Policy & $\alpha=0.2$ & $\alpha=0.5$ & $\alpha=0.8$ \\
\midrule
AUTO & 0.510101 & 0.483153 & 0.456205 \\
AUTO ($\alpha=0.2$) & 0.505359 & 0.480854 & 0.456348 \\
AUTO ($\alpha=0.5$) & 0.506077 & 0.481263 & 0.456448 \\
AUTO ($\alpha=0.8$) & 0.507218 & 0.480113 & 0.453008 \\
GDSF & 0.510865 & 0.481510 & 0.452155 \\
LHD & 0.521778 & 0.489596 & 0.457415 \\
S3-FIFO & 0.546445 & 0.517095 & 0.487745 \\
SIEVE & 0.529180 & 0.510413 & 0.491645 \\
WTinyLFU-AV & 0.552026 & 0.512398 & 0.472770 \\
\bottomrule
\end{tabular}
}
}
\end{table}

\begin{table}[htbp]
  \centering
  \caption{AUTO vs SIEVE: cacheable-only OMR vs all-request BMR tradeoffs.}
  \label{tab:tradeoffsum}
  {\setlength{\tabcolsep}{3pt}\renewcommand{\arraystretch}{1.05}% AUTO vs SIEVE: cacheable-only OMR vs all-request BMR tradeoffs.
\resizebox{\linewidth}{!}{%
\begin{tabular}{@{}>{\raggedright\arraybackslash}p{0.60\linewidth}>{\raggedleft\arraybackslash}p{0.15\linewidth}>{\raggedleft\arraybackslash}p{0.15\linewidth}@{}}
\toprule
Case & Count & Fraction \\
\midrule
OMR improves, BMR\_\allowbreak{}all worsens & 28 & 0.233 \\
OMR worsens, BMR\_\allowbreak{}all improves & 6 & 0.050 \\
Both improve & 52 & 0.433 \\
Both worsen & 34 & 0.283 \\
\bottomrule
\end{tabular}
}
}
\end{table}

\subsection{Per-workload wins}
We report win counts on cacheable-only OMR (excluding degenerate trace--size pairs where policies are indistinguishable) in the reproducibility artifact.
\textsc{AUTO} improves over SIEVE on most evaluated workloads (80/120 trace--size pairs) and beats S3-FIFO on 79/120 pairs with 22 ties.
Against S3-FIFO, it wins in large-object regimes and mostly matches it in small-object regimes.
When \textsc{AUTO} trails the best expert, a single policy typically dominates that trace across cache sizes,
suggesting that pinning the expert is appropriate once the fingerprint consistently identifies that regime.

\subsection{Trends across cache sizes}
\begin{figure}[htbp]
  \centering
  \includegraphics[width=0.96\linewidth]{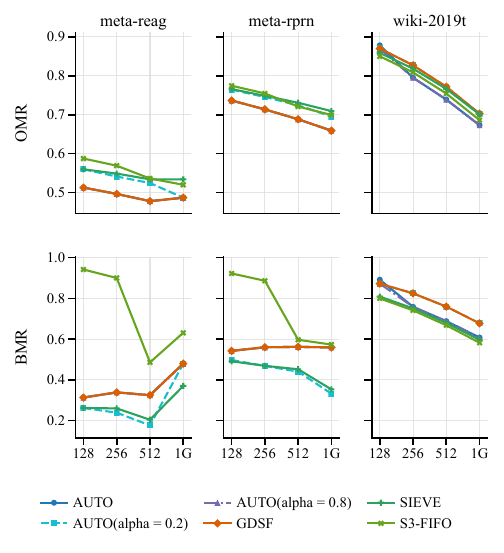}
  \caption{Cacheable-only OMR (top) and BMR (bottom) vs cache size for representative traces (lower is better for both).}
  \label{fig:omr_bmr}
\end{figure}

\Cref{fig:omr_bmr} shows OMR/BMR as cache size increases for representative traces.
The qualitative takeaway is that \textsc{SCION} avoids regime mismatch:
it selects a size-aware expert on large-object traces and selects S3-FIFO on small-object traces.

\subsection{Tail-window robustness}
We analyze worst-5\% window OMR/BMR across cache sizes as a simple tail proxy; artifact plots show that the wrong policy can incur large tail penalties.

\paragraph{Tail-aware selector.}
Using worst-5\% window OMR as the training objective (on the subset of policies with per-window stats),
we train a selector to directly target tail behavior.
Tail-aware training reduces average/P95 tail regret from 0.0190/0.098 to 0.0116/0.073 (\Cref{tab:tail_objective}),
suggesting that the fingerprint captures signal relevant to tail windows.

\begin{table}[htbp]
  \centering
  \caption{Selector performance when trained on a tail objective (non-flat samples).}
  \label{tab:tail_objective}
  {\setlength{\tabcolsep}{3pt}\renewcommand{\arraystretch}{1.05}% Selector performance on tail objective (non-flat samples).
\resizebox{\linewidth}{!}{%
\begin{tabular}{@{}>{\raggedright\arraybackslash}p{0.40\linewidth}>{\raggedleft\arraybackslash}p{0.18\linewidth}>{\raggedleft\arraybackslash}p{0.18\linewidth}>{\raggedleft\arraybackslash}p{0.18\linewidth}@{}}
\toprule
Objective & Acc $\uparrow$ & Avg regret $\downarrow$ & P95 regret $\downarrow$ \\
\midrule
OMR (subset) & 0.592 & 0.019029 & 0.098285 \\
Tail-\allowbreak{}5\% OMR & 0.625 & 0.011641 & 0.072579 \\
\bottomrule
\end{tabular}
}
}
\end{table}

\paragraph{AUTO vs fixed GDSF.}
GDSF remains the strongest average OMR baseline on this suite, while \textsc{AUTO} tends to trade small OMR losses for modest average BMR gains.
For cacheable-only BMR, \textsc{AUTO} improves over GDSF on 20\% of trace--size pairs with a mean $\Delta{=}{-}0.0024$,
aligning with the OMR/BMR tradeoff motivation.
Full AUTO--GDSF deltas (max gap 0.14 OMR; incl. worst-5\% windows) are in the reproducibility artifact.

\begin{figure}[htbp]
  \centering
  \includegraphics[width=0.94\linewidth]{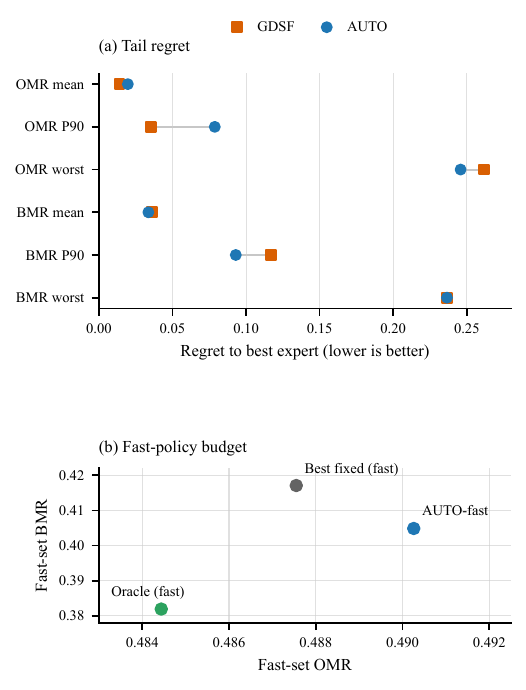}
  \caption{Two compact visual summaries of secondary analyses. Top: AUTO trades a small mean-OMR increase for better BMR tails and slightly lower worst-case OMR than fixed GDSF. Bottom: AUTO-fast improves the fast-budget Pareto point over the best fixed fast policy.}
  \label{fig:tradeoff_budget}
\end{figure}

\paragraph{Risk-sensitive regret.}
\Cref{tab:robustness_regret} reports mean, P90, and CVaR90 regret versus the best expert on cacheable-only OMR and BMR.
On OMR, GDSF has lower mean and tail regret, while \textsc{AUTO} slightly reduces the worst-case regret.
On BMR, \textsc{AUTO} lowers mean and tail regret relative to GDSF, matching the OMR/BMR tradeoff motivation.
Per-trace regret CDFs and confidence-bucket breakdowns are included in the reproducibility artifact.
Figure~\ref{fig:tradeoff_budget} gives a compact visual summary of two practical secondary results: the OMR/BMR regret tradeoff relative to fixed GDSF, and the fast-policy-budget operating point for \textsc{AUTO-fast}.

\begin{table}[htbp]
  \centering
  \caption{Risk-sensitive regret vs the best expert (cacheable-only).}
  \label{tab:robustness_regret}
  {\setlength{\tabcolsep}{3pt}\renewcommand{\arraystretch}{1.05}% Risk-sensitive regret vs the best expert (cacheable-only).
\resizebox{\linewidth}{!}{%
\begin{tabular}{@{}>{\raggedright\arraybackslash}p{0.10\linewidth}>{\raggedright\arraybackslash}p{0.18\linewidth}>{\raggedleft\arraybackslash}p{0.16\linewidth}>{\raggedleft\arraybackslash}p{0.12\linewidth}>{\raggedleft\arraybackslash}p{0.12\linewidth}>{\raggedleft\arraybackslash}p{0.12\linewidth}>{\raggedleft\arraybackslash}p{0.08\linewidth}@{}}
\toprule
Metric & Method & Mean regret & P90 & CVaR90 & Worst & N \\
\midrule
OMR & AUTO & 0.0197 & 0.0788 & 0.1265 & 0.2458 & 120 \\
OMR & GDSF & 0.0140 & 0.0351 & 0.1003 & 0.2615 & 120 \\
BMR & AUTO & 0.0336 & 0.0930 & 0.1508 & 0.2365 & 120 \\
BMR & GDSF & 0.0360 & 0.1170 & 0.1590 & 0.2365 & 120 \\
\bottomrule
\end{tabular}
}
}
\end{table}

\subsection{Sensitivity and nonstationarity}
\paragraph{Prefix stability over time.}
Our main aggregate results use a single prefix-based selection for the full run.
To quantify how stable that choice remains, we evaluate cacheable-only OMR regret per 200k-request window
relative to the per-window best expert, using the expert chosen by \textsc{AUTO} for that trace--size pair.
\Cref{tab:prefix_stability} reports post-prefix windows (excluding the selection window):
the median regret is $4.15{\times}10^{-4}$, 63.7\% of windows are within 0.01 OMR, and 82.4\% are within 0.05 OMR.
However, the prefix-best expert remains best in only 52.9\% of windows, indicating within-trace drift and motivating periodic re-selection when switching costs allow.
In spliced traces, a low-switch fixed-share baseline (alpha=0.05) matches fixed GDSF within 0.0005 OMR on average
(\Cref{tab:spliced}), suggesting that conservative re-selection can track drift without excessive oscillation.
This window-level stability provides a proxy for prefix-placement sensitivity: even if the initial prefix is atypical,
the selected expert remains close to the per-window best for most of the run.
Operationally, the p90-threshold selector is already stable with $N{=}1$k--10k requests, and $N{=}50$k performs similarly to $N{=}200$k for the multi-feature selector (\Cref{tab:short_prefix}).
We recommend choosing the smallest $N$ for which p90, cacheable fraction, and unique ratio stop changing materially across consecutive windows;
on this suite, that occurs by 1k--10k requests for \textsc{SCION-P90} and by about 50k for the multi-feature selector.
Re-select every $N$ requests (or when change-points in these statistics or low-confidence spikes appear), trading switch cost for drift tracking.

\begin{table}[htbp]
  \centering
  \caption{Prefix-based selection stability (post-prefix windows; cacheable-only OMR).}
  \label{tab:prefix_stability}
  {\setlength{\tabcolsep}{3pt}\renewcommand{\arraystretch}{1.05}% Prefix-based selection stability (post-prefix windows; cacheable-only OMR).
\resizebox{\linewidth}{!}{%
\begin{tabular}{@{}>{\raggedright\arraybackslash}p{0.62\linewidth}>{\raggedleft\arraybackslash}p{0.22\linewidth}@{}}
\toprule
Metric & Value \\
\midrule
Trace-size pairs & 98 \\
Post-prefix windows & 2282 \\
Median regret & 0.000415 \\
P90 regret & 0.091308 \\
Frac $\le 0.01$ & 0.637 \\
Frac $\le 0.02$ & 0.702 \\
Frac $\le 0.05$ & 0.824 \\
Prefix-best stays best & 0.529 \\
\bottomrule
\end{tabular}
}
}
\end{table}

\paragraph{Out-of-distribution generalization.}
We evaluate cross-family generalization by training on one dataset group and testing on another.
As a temporal split within a single dataset, training on CloudPhysics weeks 01--02 and testing on weeks 03--04 yields accuracy 0.375 with avg regret 0.0259; the reverse split is harder (avg regret 0.0679), highlighting natural drift even without cross-domain shifts.
\Cref{tab:ood} shows that training on non-Twitter traces transfers well to Twitter (acc 0.75, avg regret $6{\times}10^{-6}$),
while Twitter-only training generalizes less well to non-Twitter workloads (acc 0.47, avg regret 0.0176).
Block traces are moderately harder out-of-distribution (avg regret 0.0208 when trained on non-block traces),
underscoring the need for diverse training traces and richer fingerprints when deploying a fixed offline selector.
\paragraph{Cache-size generalization.}
Holding out one cache size at a time (train on three sizes, test on the fourth) yields 0.57--0.67 accuracy with
average regret 0.014--0.022 on the held-out size, suggesting reasonable within-range generalization (artifact table).
To test true extrapolation, we additionally ran four traces at 64\,MiB and 2\,GiB, outside the 128\,MiB--1\,GiB training range:
average regret rises to 0.029 and AUTO trails GDSF by 0.022 OMR on average, with the largest miss concentrated on CloudPhysics-w01.
In practice, deployments far outside the training range should retrain or include representative sizes in the selector's training set.

\begin{table}[htbp]
  \centering
  \caption{Out-of-distribution generalization by dataset family (OMR objective).}
  \label{tab:ood}
  {\setlength{\tabcolsep}{3pt}\renewcommand{\arraystretch}{1.05}% Out-of-distribution generalization by dataset group (OMR objective).
\resizebox{\linewidth}{!}{%
\begin{tabular}{@{}>{\raggedright\arraybackslash}p{0.40\linewidth}>{\raggedleft\arraybackslash}p{0.12\linewidth}>{\raggedleft\arraybackslash}p{0.18\linewidth}>{\raggedleft\arraybackslash}p{0.18\linewidth}>{\raggedleft\arraybackslash}p{0.08\linewidth}@{}}
\toprule
Train -> Test & Acc & Avg regret & P95 regret & N \\
\midrule
non-twitter -> twitter & 0.750 & 0.000006 & 0.000023 & 24 \\
twitter -> non-twitter & 0.469 & 0.017631 & 0.078119 & 96 \\
non-block -> block & 0.521 & 0.020802 & 0.101020 & 48 \\
block -> non-block & 0.319 & 0.008301 & 0.041023 & 72 \\
cloudphysics early -> cloudphysics late & 0.375 & 0.025894 & 0.053132 & 8 \\
cloudphysics late -> cloudphysics early & 0.375 & 0.067903 & 0.147991 & 8 \\
\bottomrule
\end{tabular}
}
}
\end{table}

\Cref{tab:sensitivity} reports a compact sensitivity summary over $N$ and $\tau$, including alternative fingerprints.
We find that percentile-based fingerprints (p90) reduce regret relative to the mean under heavy-tailed sizes.
Notably, the best p90 setting already uses $N{=}1$k--10k requests, indicating that very short prefixes can be sufficient for selection.
\Cref{tab:short_prefix} summarizes p90-threshold performance across prefix lengths (best $\tau$ per $N$);
regret is essentially unchanged from $N{=}1$k to $N{=}200$k on this trace suite, suggesting that very short prefixes
can drive the cheap threshold selector; at $N{=}50$k the regret matches the $N{=}200$k setting (\Cref{tab:short_prefix}).
\Cref{tab:spliced} evaluates online epochal selection on spliced traces (large$\rightarrow$small and small$\rightarrow$large),
including hysteresis (lag/min-stay) and window-level bandit baselines (epsilon-greedy, Hedge, and EXP3; LeCaR-style online expert tracking);
infrequent re-selection can track regime changes with limited degradation.
These spliced experiments switch experts without state transfer (cold-start), which upper-bounds practical switching cost
when warm transitions or shadowing are used in systems; our prototype warm-switch recovers up to 3 percentage points of hit ratio (artifact).
We additionally include a fixed-share (tracking) meta-algorithm with switching-cost awareness and an offline switching-cost oracle
to contextualize online performance under regime changes.
We model switching cost as an additive 2\% miss penalty per switch in these simulations (switch counts are summarized in the reproducibility artifact).
These serve as empirical reference points; we do not derive formal tracking or switching-cost guarantees in this work.
The full per-trace spliced results are in the reproducibility artifact.
Average switch counts show that the threshold and fixed-share selectors switch only a handful of times per run,
while bandit/EXP3-style baselines switch far more frequently.

\begin{table}[htbp]
  \centering
  \caption{Fingerprint sensitivity (best setting per fingerprint across $N$ and $\tau$).}
  \label{tab:sensitivity}
  {\setlength{\tabcolsep}{3pt}\renewcommand{\arraystretch}{1.05}% Best (lowest avg regret) setting per fingerprint across $N$ and $\tau$.
\resizebox{\linewidth}{!}{%
\begin{tabular}{@{}>{\raggedright\arraybackslash}p{0.20\linewidth}>{\raggedleft\arraybackslash}p{0.10\linewidth}>{\raggedleft\arraybackslash}p{0.20\linewidth}>{\raggedleft\arraybackslash}p{0.15\linewidth}>{\raggedleft\arraybackslash}p{0.15\linewidth}>{\raggedleft\arraybackslash}p{0.15\linewidth}@{}}
\toprule
Fingerprint & $N$ & $\tau$ (bytes) & Miscls & Avg regret & Worst regret \\
\midrule
logmean & 1000 & 262144 & 0.742 & 0.036269 & 0.542588 \\
mean & 1000 & 262144 & 0.608 & 0.031072 & 0.542588 \\
median & 1000 & 262144 & 0.708 & 0.036070 & 0.542588 \\
p90 & 1000 & 262144 & 0.475 & 0.025470 & 0.542588 \\
trimmed & 5000 & 262144 & 0.608 & 0.031072 & 0.542588 \\
\bottomrule
\end{tabular}
}
}
\end{table}

\begin{table}[htbp]
  \centering
  \caption{Short-prefix p90 threshold sensitivity (best $\tau$ per $N$).}
  \label{tab:short_prefix}
  {\setlength{\tabcolsep}{3pt}\renewcommand{\arraystretch}{1.05}% Short-prefix sensitivity for p90 fingerprint (best tau per N).
\resizebox{\linewidth}{!}{%
\begin{tabular}{@{}>{\raggedright\arraybackslash}p{0.25\linewidth}>{\raggedright\arraybackslash}p{0.20\linewidth}>{\raggedleft\arraybackslash}p{0.22\linewidth}>{\raggedleft\arraybackslash}p{0.22\linewidth}@{}}
\toprule
Prefix $N$ & Best $\tau$ & Avg regret $\downarrow$ & P95 regret $\downarrow$ \\
\midrule
1K & 256KiB & 0.025470 & 0.029927 \\
5K & 256KiB & 0.026814 & 0.031231 \\
10K & 256KiB & 0.026814 & 0.031231 \\
50K & 256KiB & 0.026814 & 0.031231 \\
200K & 256KiB & 0.026814 & 0.031231 \\
1M & 256KiB & 0.027014 & 0.031437 \\
\bottomrule
\end{tabular}
}
}
\end{table}

\begin{table}[htbp]
  \centering
  \caption{Online epochal selection on spliced traces (summary; cacheable-only OMR).}
  \label{tab:spliced}
  {\setlength{\tabcolsep}{3pt}\renewcommand{\arraystretch}{1.05}% Summary of spliced-trace results (lower is better).
\resizebox{\linewidth}{!}{%
\begin{tabular}{@{}>{\raggedright\arraybackslash}p{0.42\linewidth}>{\raggedleft\arraybackslash}p{0.18\linewidth}>{\raggedleft\arraybackslash}p{0.18\linewidth}>{\raggedleft\arraybackslash}p{0.18\linewidth}@{}}
\toprule
Mode & Avg OMR $\downarrow$ & Best OMR $\downarrow$ & Worst OMR $\downarrow$ \\
\midrule
oracle\_\allowbreak{}switch(cost=0.\allowbreak{}02) & 0.566952 & 0.477694 & 0.647771 \\
fixed\_\allowbreak{}share(alpha=0.\allowbreak{}05) & 0.569891 & 0.478663 & 0.652027 \\
fixed:GDSF & 0.570428 & 0.477694 & 0.654674 \\
hedge & 0.571376 & 0.478663 & 0.655083 \\
threshold & 0.571516 & 0.479902 & 0.654131 \\
bandit & 0.587959 & 0.494979 & 0.694912 \\
fixed:SIEVE & 0.595822 & 0.500604 & 0.676335 \\
exp3 & 0.604616 & 0.505300 & 0.686499 \\
fixed:S3-FIFO & 0.607712 & 0.504683 & 0.691214 \\
\bottomrule
\end{tabular}
}
}
\end{table}

\subsection{Synthetic stress tests}
To probe extreme regimes not well represented in public traces, we generate synthetic oracleGeneral traces
with (i) very high hotspots (Zipf $\theta{=}1.1$), (ii) high-churn/near-write-dominated streams (95\% unique ratio),
and (iii) rapid regime switches (hotspot$\rightarrow$uniform).
We report cacheable-only OMR/BMR at 256\,MiB over 1M requests per synthetic trace.
\Cref{tab:synthetic} (cells show OMR/BMR) indicates that the same regime split persists: size-aware policies dominate large-object regimes,
while FIFO-family policies dominate small-object regimes; \textsc{SCION-P90} avoids the worst mismatches.

\begin{table}[htbp]
  \centering
  \caption{Synthetic stress tests (cacheable-only OMR/BMR at 256\,MiB, 1M requests).}
  \label{tab:synthetic}
  {\setlength{\tabcolsep}{3pt}\renewcommand{\arraystretch}{1.05}\resizebox{\linewidth}{!}{%
\begin{tabular}{@{}lccccc@{}}
\toprule
Trace & S3-FIFO & GDSF & SIEVE & ARC & SCION-P90 \\
\midrule
Hotspot (Zipf 1.1), large objects & 0.472/0.470 & 0.389/0.388 & 0.508/0.507 & 0.492/0.490 & 0.389/0.388 \\
Hotspot (Zipf 1.1), small objects & 0.155/0.154 & 0.136/0.135 & 0.155/0.154 & 0.150/0.149 & 0.155/0.154 \\
High churn (95% unique) & 1.000/1.000 & 1.000/1.000 & 1.000/1.000 & 1.000/1.000 & 1.000/1.000 \\
Regime switch (hotspot->uniform) & 0.668/0.475 & 0.527/0.393 & 0.740/0.507 & 0.683/0.525 & 0.527/0.393 \\
\bottomrule
\end{tabular}
}
}
\end{table}

\subsection{Overhead-aware budget}
To connect throughput overheads to operator decisions, we define a fast-policy budget using median MQPS
from \Cref{tab:overhead}: policies with $\mathrm{MQPS} \ge 4.5$ (SIEVE, S3-FIFO, LHD, AdaptiveClimb, DynamicAdaptiveClimb).
We retrain the selector on this restricted set (AUTO-fast) and compare against the best fixed fast policy and the per-trace oracle.
AUTO-fast is trained on the same $\alpha{=}0.5$ cost objective used in \Cref{tab:cost} (within the fast set).
\Cref{tab:overhead_budget} shows that AUTO-fast trades a small OMR increase for a larger BMR reduction; under the $\alpha{=}0.5$ cost
objective used for training, this yields a lower average cost than the best fixed fast policy (LHD).
The oracle row quantifies remaining headroom. This illustrates how SCION can incorporate CPU budgets by constraining or reweighting
the expert set.
Because these budgets depend on \texttt{libCacheSim} implementations, production-grade TinyLFU variants (e.g., Caffeine/Ristretto)
could shift the fast set; we did not benchmark them here, but the same MQPS-driven filter would apply.
Accordingly, the 4.5 MQPS threshold should be read as a simulator-specific screening rule, not a universal deployment constant.
Given \Cref{tab:overhead}, WTinyLFU-AV would require roughly a 100$\times$ speedup to meet the 4.5 MQPS fast-set threshold.
As a sensitivity check, if WTinyLFU-AV were fast enough to qualify, the fast-set oracle would improve by $\approx$0.008 OMR and $\approx$0.014 BMR
(cacheable-only averages); we do not retrain AUTO-fast under this assumption.

\begin{table}[htbp]
  \centering
  \caption{Overhead-aware selection under a fast-policy budget (cacheable-only averages).}
  \label{tab:overhead_budget}
  {\setlength{\tabcolsep}{3pt}\renewcommand{\arraystretch}{1.05}% Overhead-aware selection with a fast-policy budget (cacheable-only).
\resizebox{\linewidth}{!}{%
\begin{tabular}{@{}>{\raggedright\arraybackslash}p{0.32\linewidth}>{\raggedright\arraybackslash}p{0.24\linewidth}>{\raggedleft\arraybackslash}p{0.18\linewidth}>{\raggedleft\arraybackslash}p{0.18\linewidth}@{}}
\toprule
Selector & Policy & Avg OMR $\downarrow$ & Avg BMR $\downarrow$ \\
\midrule
Best fixed (fast) & LHD & 0.487550 & 0.417048 \\
AUTO-fast & AUTO-fast & 0.490259 & 0.404845 \\
Oracle (fast) & oracle & 0.484438 & 0.381937 \\
\bottomrule
\end{tabular}
}
}
\end{table}

\subsection{Overheads}
\paragraph{Fingerprint cost.}
Computing the 200k-request prefix fingerprint takes about 1.6\,s on our CPU-only setup (median across traces),
and runs once per epoch/window rather than on the hot path.
Across 10 smaller traces (1M requests, 200k-request windows), the median per-window fingerprint cost is 1.64\,s (p90 1.70\,s),
equivalent to $\approx$8\,\textmu s per request when amortized, or $\approx$8.2\,s of background CPU time per 1M requests.
We provide a script to reproduce these cumulative epoch-level measurements.
\paragraph{Per-policy throughput.}
\Cref{tab:overhead} reports single-threaded throughput for representative policies.
Queue-based FIFO-family policies are fastest; GDSF is slower due to priority-queue updates, and W-TinyLFU-AV
incurs substantial overhead from its admission sketch. AdaptiveClimb/DynamicAdaptiveClimb are comparable to SIEVE.
S3-FIFO in \texttt{libCacheSim} maintains multiple queues and a ghost list and is not heavily optimized, which can depress throughput;
production implementations may differ.
Our W-TinyLFU-AV implementation is the \texttt{libCacheSim} research version; production-grade TinyLFU deployments
(e.g., Caffeine, Ristretto) are heavily optimized and can have materially higher throughput~\cite{caffeine,ristretto}.
Accordingly, we treat MQPS as an implementation-level overhead indicator rather than an algorithmic guarantee and avoid
over-interpreting cross-policy throughput rankings.
We do not yet proxy production admission paths in our harness; integrating those variants is future work that could refine
the fast-policy budget results.

\begin{table}[htbp]
  \centering
  \caption{Per-policy throughput microbenchmark.}
  \label{tab:overhead}
  {\setlength{\tabcolsep}{3pt}\renewcommand{\arraystretch}{1.05}% Single-threaded policy throughput (2M requests, 6 traces).
\resizebox{\linewidth}{!}{%
\begin{tabular}{@{}>{\raggedright\arraybackslash}p{0.44\linewidth}>{\raggedleft\arraybackslash}p{0.23\linewidth}>{\raggedleft\arraybackslash}p{0.23\linewidth}@{}}
\toprule
Policy & Avg MQPS & Rel. to SIEVE \\
\midrule
SIEVE & 8.520 & 1.00x \\
AdaptiveClimb & 8.512 & 1.00x \\
DynamicAdaptiveClimb & 8.188 & 0.96x \\
S3-FIFO & 5.708 & 0.67x \\
LHD & 4.945 & 0.58x \\
GDSF & 2.738 & 0.32x \\
WTinyLFU-AV & 0.047 & 0.01x \\
\bottomrule
\end{tabular}
}
}
\end{table}

\subsection{End-to-end prototype}
\Cref{tab:systemproto} summarizes the real-system prototype using 20k-request prefixes per trace.
Even with a lightweight implementation, the same regime split is visible:
\textsc{AUTO} selects the appropriate expert and reduces origin traffic while maintaining good latency and throughput, including against SIEVE.
The prototype uses \textsc{SCION-P90} (no offline training); in the full simulator, \textsc{SCION-P90} trails \textsc{AUTO} by
0.0205 OMR and 0.0062 BMR on average (artifact table), so the prototype remains representative of the learned selector's regime choices.

\begin{table*}[htbp]
  \centering
  \caption{End-to-end prototype results (cache-daemon + origin).}
  \label{tab:systemproto}
  {\setlength{\tabcolsep}{3pt}\renewcommand{\arraystretch}{1.02}% End-to-end prototype results (cache-daemon + origin).
\resizebox{\linewidth}{!}{%
\begin{tabular}{@{}>{\raggedright\arraybackslash}p{0.26\linewidth}>{\raggedright\arraybackslash}p{0.10\linewidth}>{\raggedright\arraybackslash}p{0.06\linewidth}>{\raggedleft\arraybackslash}p{0.06\linewidth}>{\raggedleft\arraybackslash}p{0.06\linewidth}>{\raggedleft\arraybackslash}p{0.06\linewidth}>{\raggedleft\arraybackslash}p{0.06\linewidth}>{\raggedleft\arraybackslash}p{0.06\linewidth}>{\raggedleft\arraybackslash}p{0.06\linewidth}@{}}
\toprule
Trace & Policy & Cache & P50 ms & P95 ms & P99 ms & RPS & Hit & Origin GB \\
\midrule
meta\_\allowbreak{}reag.\allowbreak{}oracleGeneral.\allowbreak{}zst & SIEVE & 256MiB & 46.62 & 3023.18 & 8214.08 & 128.5 & 0.319 & 290.87 \\
meta\_\allowbreak{}reag.\allowbreak{}oracleGeneral.\allowbreak{}zst & GDSF & 256MiB & 41.96 & 3035.64 & 8774.38 & 128.8 & 0.365 & 298.09 \\
meta\_\allowbreak{}reag.\allowbreak{}oracleGeneral.\allowbreak{}zst & AUTO & 256MiB & 43.74 & 3162.16 & 8368.71 & 123.6 & 0.364 & 296.03 \\
meta\_\allowbreak{}reag.\allowbreak{}oracleGeneral.\allowbreak{}zst & S3-FIFO & 256MiB & 48.36 & 3038.00 & 7871.37 & 120.2 & 0.317 & 291.14 \\
meta\_\allowbreak{}rprn.\allowbreak{}oracleGeneral.\allowbreak{}zst & GDSF & 256MiB & 53.49 & 1358.30 & 4809.51 & 73.0 & 0.240 & 648.03 \\
meta\_\allowbreak{}rprn.\allowbreak{}oracleGeneral.\allowbreak{}zst & S3-FIFO & 256MiB & 46.94 & 1170.62 & 3870.59 & 71.7 & 0.206 & 647.41 \\
meta\_\allowbreak{}rprn.\allowbreak{}oracleGeneral.\allowbreak{}zst & SIEVE & 256MiB & 48.39 & 1212.44 & 4285.68 & 74.1 & 0.206 & 647.65 \\
meta\_\allowbreak{}rprn.\allowbreak{}oracleGeneral.\allowbreak{}zst & AUTO & 256MiB & 47.34 & 1233.02 & 4134.08 & 73.5 & 0.239 & 647.92 \\
wiki\_\allowbreak{}2019t.\allowbreak{}oracleGeneral.\allowbreak{}zst & GDSF & 256MiB & 9.39 & 11.23 & 14.31 & 13598.2 & 0.119 & 0.59 \\
wiki\_\allowbreak{}2019t.\allowbreak{}oracleGeneral.\allowbreak{}zst & S3-FIFO & 256MiB & 10.47 & 29.04 & 31.18 & 8376.1 & 0.114 & 0.58 \\
wiki\_\allowbreak{}2019t.\allowbreak{}oracleGeneral.\allowbreak{}zst & SIEVE & 256MiB & 9.00 & 10.39 & 16.24 & 14220.0 & 0.111 & 0.59 \\
wiki\_\allowbreak{}2019t.\allowbreak{}oracleGeneral.\allowbreak{}zst & AUTO & 256MiB & 10.59 & 28.66 & 31.03 & 8439.4 & 0.114 & 0.58 \\
\bottomrule
\end{tabular}
}
}
\end{table*}

\paragraph{Switching-cost microbenchmark.}
We approximate the operational cost of switching experts by splitting a 20k-request trace into two halves and comparing
\emph{warm} (shadow-cache) versus cold switches.
A warm switch recovers up to 3 percentage points of hit ratio versus cold-start (details in the artifact),
so the spliced-trace results that assume cold-start are conservative for deployments that can warm transitions.

\section{Discussion and Outlook}
\label{sec:discussion}
\paragraph{Interpretation.}
The result is not that a new policy beats all baselines everywhere; rather, the orchestration layer helps avoid regime mismatches
when the operator objective is multi-dimensional (OMR vs BMR) or subject to CPU budgets.
In workloads where a baseline is already optimal (e.g., small-object regimes for S3-FIFO), the right behavior is to select it.
This tradeoff- and overhead-focused goal aligns with the direction of recent learned caching work~\cite{halp,mat,threeLcache}.
Accordingly, \textsc{AUTO} is positioned as a pragmatic default for heterogeneous deployments, not as a universal winner on any single average metric.
We do not introduce a new replacement algorithm; learning is used only to select among existing experts off the hot path.
Practically, CloudPhysics-like traces and any deployment with persistent high fallback or a single-expert winner across sizes are ``do not orchestrate'' cases where pinning GDSF/S3-FIFO is preferable.

\paragraph{What this does \emph{not} solve yet.}
Our prototype uses a small expert set and a tiny feature vector, but still stops short of full online selection with delayed feedback
and does not yet optimize tail-latency proxies.
The end-to-end prototype is intentionally minimal: it omits TTLs, revalidation, invalidations, and distributed consistency,
which are important in production caches but orthogonal to the selection mechanism studied here.
We also do not yet report full-trace results for every classic or cost-aware policy (e.g., ARC/LIRS/TwoQ, CAMP) or production-grade
TinyLFU implementations across all sizes;
the harness supports many of these, and expanding that coverage is straightforward but compute-intensive.
Our main sweep covers 128--1024\,MiB. The 64\,MiB/2\,GiB spot-check above suggests within-range generalization is materially
better than true extrapolation, so deployments far outside that range should retrain or include those sizes; broader normalized
$C$/working-set analyses remain future work.
Finally, our confidence score is a coarse margin; it is not calibrated, and fallback defaults to GDSF in our experiments.
\paragraph{Switching costs (prototype view).}
Our prefix-stability analysis shows that some degree of drift is common even when median regret is small.
The microbenchmark above suggests that simple shadow-cache warmup can reduce switching penalties; full production costs remain future work.

\section{Related Work}
\label{sec:related}
\paragraph{Learned caching for systems.}
LRB~\cite{lrb} and Cacheus~\cite{cacheus} explore learned admission/replacement and mixtures of experts;
Cacheus maintains per-request mixture weights, whereas \textsc{SCION} selects per-trace/epoch based on a tiny fingerprint.
This decoupling enables explicit reject-option fallback and throughput budgeting without per-request ML.
Recent work emphasizes deployability and low overhead (e.g., HALP, MAT, and \texttt{3L-Cache})~\cite{halp,mat,threeLcache}.
We compare to Cacheus/LeCaR on a small subset due to harness cost; HALP/MAT/\texttt{3L-Cache} implementations are not available in \texttt{libCacheSim}
and place ML on the hot path, so we treat them as complementary and discuss their overhead relative to our fast-policy budget.
Unlike these methods, \textsc{SCION} keeps inference entirely off the hot path and only selects among existing experts,
trading potential algorithmic optimality for operational simplicity and robustness.
HR-Cache~\cite{hrcache}, Cold-RL~\cite{coldrl}, and LCR/LARU~\cite{lcr_laru} instead keep learning on the hot path while targeting BMR or bounded overhead.
\paragraph{TTL-based admission and expiration.}
TTL policies control admission and eviction using per-object time-to-live values, with adaptive variants such as d-TTL and f-TTL
that tune TTLs to hit-rate or cache-size objectives under nonstationary demand~\cite{adaptive_ttl}.
These methods are complementary to \textsc{SCION}; TTL rules can act as experts or as admission filters upstream of replacement.
\paragraph{Online selection and no-regret caching.}
Regret-minimizing approaches such as LeCaR~\cite{lecar} and recent online gradient-based caching with logarithmic complexity~\cite{ogb}
provide theoretical guarantees under adversarial or nonstationary sequences.
\textsc{SCION} targets a different point in the design space: infrequent, off-path selection among a small expert set with minimal runtime state.
For online expert tracking, fixed-share and related algorithms provide regret bounds against the best switching sequence~\cite{herbster1998tracking,cesa2006prediction};
we include such baselines in our spliced-trace analysis.
\paragraph{Offline bounds and headroom.}
For variable-size caching, Belady is only one offline reference point.
FOO/PFOO formulate tighter offline bounds via min-cost flow and quantify remaining OPT headroom under variable object sizes~\cite{foo_pfoo}.
Our Belady/BeladySize subset should therefore be read as a lightweight oracle sanity check, not as the last word on offline headroom.

\paragraph{Learning for replacement (broader contexts).}
Imitation-learning and ``approximate MIN'' approaches (e.g., Parrot; Hawkeye/OPTgen) aim to mimic Belady-style decisions~\cite{parrot,hawkeye}.
Other learned replacement work explores RL-derived policies and reuse-distance prediction (e.g., RLR, Mockingjay)~\cite{rlr,mockingjay}.
These lines inform \textsc{SCION}'s ethos: use learning to \emph{select} or \emph{approximate} what is hard, while keeping the online path simple.

\paragraph{Strong non-ML object caching.}
SIEVE and S3-FIFO are recent strong baselines~\cite{sieve,s3fifo}; classic size-aware scoring (GDS/GDSF) remains effective in large-object regimes~\cite{cao1997costaware},
and benefit/density-based heuristics such as LHD have been studied for object caches~\cite{lhd}.
AdaptiveClimb and DynamicAdaptiveClimb offer low-overhead alternatives with strong performance across diverse traces~\cite{dynamicadaptiveclimb}.
Clock2Q+ extends queue-based designs for metadata caches and reports strong low-overhead performance~\cite{clock2qplus};
its target workload is storage metadata rather than variable-size CDN/object traces, so we treat it as a future expert candidate rather than a direct baseline here.
Cost-aware approximations to GDS and production-grade TinyLFU variants (e.g., Caffeine/Ristretto-style implementations)
are natural expert candidates for \textsc{SCION}~\cite{caffeine,ristretto}.

\section{Conclusion}
\label{sec:concl}
We presented \textsc{SCION}, a lightweight orchestration framework for object-cache replacement policies.
The main result is not that \textsc{AUTO} dominates every expert on average; it is that a tiny off-path fingerprint is often enough to automate policy choice,
expose explicit OMR/BMR/CPU tradeoffs, and identify when operators should pin a single expert instead.
Across public traces and synthetic stress tests, the main regime split is size-aware vs FIFO-family, while the remaining oracle gap is more pronounced for byte-aware objectives than for OMR.
We provide a reproducible benchmark, selector training pipeline, and prototype; future work will add richer offline bounds, production-grade fast baselines, and fuller online switching.

\balance
\bibliographystyle{abbrv}
\bibliography{references}

\end{document}